\documentclass[iop]{emulateapj}

\providecommand{\putpic}[2][1]{\includegraphics[width=#1\linewidth]{#2}}
\providecommand{\nopic}[1]{{\color{red} Image: \detokenize{#1} not found!}}
\providecommand{\pic}[2][1]{ \IfFileExists{#2.pdf}{\putpic[#1]{#2}}{\IfFileExists{#2.png}{ \putpic[#1]{#2} } { \IfFileExists{#2.jpg}{\putpic[#1]{#2}}{\nopic{#2}} } } }

\usepackage{amsmath, amsthm, amssymb,amsfonts}
\usepackage{url}

\usepackage{tikz}
\usetikzlibrary{shapes, arrows, shadows}
\usetikzlibrary{decorations.pathreplacing}

\providecommand{\pic}[2][1]{\includegraphics[width=#1\linewidth]{pics/#2}}

\providecommand{\nolinkurl}[1]{\url{#1}}

\usepackage{lipsum}

\newcommand{\picsize}{1}

\begin{document}
	
	\title{The Weizmann Fast Astronomical Survey Telescope (W-FAST): System Overview}
	
	\author{Guy Nir\altaffilmark{1}}
	\author{Eran O.~Ofek\altaffilmark{1}}
	\author{Sagi Ben-Ami\altaffilmark{1}}
	\author{Noam Segev\altaffilmark{1}}
	\author{David Polishook\altaffilmark{1}}
	\author{Ofir Hershko\altaffilmark{1}}
	\author{Oz Diner\altaffilmark{1}}
	\author{Ilan Manulis\altaffilmark{1}}
	\author{Barak Zackay\altaffilmark{1}}
	\author{Avishay Gal-Yam\altaffilmark{1}}
	\author{Ofer Yaron\altaffilmark{1}}
	
	\altaffiltext{1}{Benoziyo Center for Astrophysics, Weizmann Institute
		of Science, 76100 Rehovot, Israel}
	
	\begin{abstract}
		A relatively unexplored phase space of transients and stellar variability
		is that of second and sub-second time-scales. 
		We describe a new optical observatory operating in the Negev desert in Israel, 
		with a 55 cm aperture, a field of view of $2.6\times 2.6^\circ$ ($\approx 7$\,deg$^2$)
		equipped with a high frame rate, low read noise, CMOS camera. 
		The system can observe at a frame rate of up to 90\,HZ (full frame), 
		while nominally observations are conducted at 10--25\,Hz. 
		The data, generated at a rate of over 6 Gbits/s at a frame rate of 25\,Hz, 
		is analyzed in real time. 
		The observatory is fully robotic and
		capable of autonomously collecting data on a few thousand stars in each field each night. 
		We present the system overview, 
		performance metrics, 
		science objectives, 
		and some first results, 
		e.g., the detection of a high rate of glints
		from geosynchronous satellites, 
		reported in Nir et al.~2020. 
		
	\end{abstract}
	
	
	\section{Introduction}
	
		Traditionally, the transient and variable sky are explored on an hours-to-weeks time scale
		(e.g., \citealt{survey_telescope_SDSS_Gunn_2006,large_synoptic_survey_telescope_Ivezic_2007,palomar_transient_factory_Law_Kulkarni_2009,catalina_real_time_transient_survey_Djorgovski_2010,PanSTARRS_survey_Kaiser_2010,catalina_sky_survey_Christensen_2012,Zwicky_transient_facility_Bellm_Kulkarni_2019}), 
		while shorter cadence (e.g., sub-seconds to minutes)
		are less common, usually limited to small fields of view,
		single sources, or specific science cases (e.g., exoplanets, \citealt{kepler_mission_Borucki_2007,transiting_exoplanet_survey_satellite_Ricker_2014}).
		In order to find new phenomena and characterize some
		known rare objects, there is a need to 
		increase both the sampling rate and the field of view (i.e., increasing telescope grasp; e.g., \citealt{small_telescopes_Ofek_Ben_Ami_2020}). 
		These requirements are typically limited by technology,
		and the cost of current instrumentation.
		With the advance of high-quality large format CMOS devices it is now becoming possible to explore the sky on sub-second time scales with a large field of view. 
		As a result, an increasing number of sky surveys with sub-second imaging capabilities are being constructed
		(e.g., \citealt{TAOS_survey_Zhang_2013,TAOS_II_camera_Wang_2016,Colibri_survey_Pass_2018,KBO_detection_from_ground_Arimatsu_2019}).
		
		There are several known phenomena which are expected to have time scales below a few minutes in the visible light band.
		These include stellar-mass accreting black holes \citep{galactic_x_ray_sources_INTEGRAL_Krivonos_2012,x_ray_binaries_black_hole_review_Remillard_McClintock_2006},
		flare stars \citep{flare_stars_Balona_2015,flare_stars_West_2015}, 
		fast rotating white dwarfs \citep{intermediate_polars_review_Patterson_1994,cataclysmic_variables_intermediate_polars_space_density_Pretorius_Mukai_2014}, 
		super-fast rotation of small-sized asteroids (e.g., \citealt{fast_rotating_asteroids_Polishook_2012,fast_rotating_asteroids_Thirouin_2016}),
		and occultation by Solar System objects such as Kuiper Belt Objects (KBOs) and Oort cloud objects
		(e.g., \citealt{occultation_idea_Bailey_1976,KBO_occultation_diffraction_Roques_1987,KBO_single_object_Schlichting_Ofek_2009,abundance_kuiper_objects_Schlichting_Ofek_2012,KBO_detection_from_ground_Arimatsu_2019}).
		Other fast phenomena may be revealed as we explore the new parameter space 
		of high cadence and wide field of view. 
		Examples for such events are a galactic GRB discussed in \cite{x_ray_binary_GRB_transient_Kasliwal_2008}
		and possible counterparts to Fast Radio Bursts (FRBs; \citealt{FRB_discovery_Lorimer_2007,ASKAP_science_overview_2008,CHIME_system_overview_2018,SGR1935_x_ray_infrared_flares_De_2020}). 
		Recently, \cite{satellite_glints_EvryScope_Corbett_2020} and \cite{satellite_glints_WFAST_Nir_2020}
		reported the detection of a high rate of sub-second transients  
		originating from reflections of sunlight by geosynchronous satellites. 
		Such glints would become a dominant foreground for 
		upcoming sky surveys like the Legacy Survey of Space and Time (LSST; \citealt{large_synoptic_survey_telescope_Ivezic_2007}). 
		
		Given these science cases, several instruments were built, or are being constructed, for monitoring
		a large field of view with sub-second cadence.
		A list of some of the instruments that provide sub-second time resolution for more than a few stars simultaneously is shown in Table~\ref{tab: other surveys}.
		
		\begin{deluxetable*} {llclllll} 
			\tabletypesize{\footnotesize}
			\tablewidth{0pt}
			
			\tablecaption{Comparison of W-FAST with other high-cadence, wide-field surveys.}
			
			\tablehead{
				\colhead{Survey} & \colhead{Telescope statistics} & \colhead{$N_\text{tel}$} & \colhead{F.O.V (deg$^2$)} & \colhead{Rate (Hz) } & \colhead{Location} &\colhead{Survey start} & \colhead{Reference}
			}

		 	\startdata 
		 		W-FAST                       &   55\,cm f/1.9     &  1      & 7                   &   10--25     & Israel   &  2019   & This work \\ 
				AOSES                        &   28\,cm f/1.58    &  2      & 4                   &   15.4       & Japan    &  2017   & \cite{OASES_survey_Arimatsu_2017} \\
				TAOS-I                       &   50\,cm f/1.9     &  4      & 3                   &   5          & Taiwan   &  2005   & \cite{TAOS_survey_Zhang_2013} \\
				TAOS-II                      &   1.3\,m f/4       &  (3)    & 1.7                 &   20         & Mexico   &  2021?  & \cite{TAOS_II_camera_Wang_2016} \\ 
				Colibri                      &   50\,cm f/2.2     &  3      & 0.47                &   40         & Canada   &  2021?  & \cite{Colibri_survey_Pass_2018} \\ 
				Tomoe Gozen                  &   1\,m f/3.1       &  1      & 20.7                &   2          & Japan    &  2019   & \cite{Tomoe_Gozen_camera_Sako_2018} \\ 
				CHIMERA                      &   5.1\,m f/3.5     &  1      & $1.4\cdot 10^{-3}$  &   40         & USA      &  2016   & \cite{Chimera_fast_camera_Harding_Hallinan_2016} \\
				ULTRACAM on WHT              &   4.2\,m f/2.1     &  1      & $7 \cdot 10^{-3}$   &   500        & La Palma &  2001   & \cite{ULTRACAM_system_Dhillon_2007}
			\enddata

		 
		   \label{tab: other surveys}
		 
		\end{deluxetable*}
		
		The Weizmann Fast Astronomical Survey Telescope (W-FAST), 
		presented in this work, 
		operates a 55\,cm and f/1.9 Schmidt telescope
		equipped with a fast readout sCMOS-camera 
		with a field of view of $\approx 7$\,deg$^2$. 
		The telescope typically observes a few thousand stars
		at a cadence of 10--25\,Hz. 
		
		The W-FAST science goals include exploring the variable and transient sky 
		at sub-minute time scales, 
		and searching for occultations of field stars by trans-Neptunian objects. 
		For these purposes, 
		W-FAST produces lightcurves and raw-data cutouts for 
		a few thousand of the brightest stars in the field at a cadence of 10--25\,Hz, 
		as well as full-frame coadded images with a cadence of 4--10 seconds.
		The coadded images provide measurements of dimmer, lower-time-scale objects. 
		These deeper coadds are used to detect or follow-up 
		M-dwarf flares,
		close binaries, 
		variability of accreting compact objects, 
		FRB optical counterparts,
		Near Earth Objects (NEOs), 
		and more (see Section~\ref{sec: science}). 
		
		We give an overview of the system in Section~\ref{sec: system overview}, 
		we describe the observatory control system in Section~\ref{sec: observatory control system}, 
		and the data acquisition and processing in Section~\ref{sec: data acquisition and processing}. 
		We describe some of the main science goals of W-FAST in Section~\ref{sec: science} 
		and show some preliminary results in Section~\ref{sec: first results}. 
		We conclude with a short summary in Section~\ref{sec: summary}.

	\section{System overview}\label{sec: system overview}
		
%
		W-FAST is a 55\,cm f/1.9 Schmidt. 
		The corrected area of the focal plane, 
		providing theoretical image quality better than about $1''$, 
		is up to $9\times 9$\,cm, 
		corresponding to a field of view of about 23\,deg$^2$.
		Our current camera has a sensor size of $5\times5$\,cm
		with a field of view of about 7\,deg$^{2}$.
		A summary of the observatory and telescope properties are listed in Table~\ref{tab: properties}.
		
		\begin{deluxetable}{lr}
			
			\tabletypesize{\footnotesize}
			\tablewidth{0pt}
			\tablecaption{Summary of the properties of the W-FAST system.}

			\tablehead{
				\colhead{Property} & \colhead{Value} 
			}
			
			\startdata 
				design                                         & Schmidt             \\
				aperture                                       & 55\,cm              \\
				focal length                                   & 108\,cm             \\ 
				f-number                                       & f/1.96              \\ 
				mount type                                     & German-Equatorial   \\	\hline 
				\rule{0pt}{2.6ex}camera                        & Andor Balor         \\
				sensor type                                    & sCMOS               \\
				sensor size                                    & $5\times 5$\,cm     \\
				pixel size                                     & 12.2\,$\mu$m        \\
				number of pixels                               & $4128\times 4104$   \\
				field of view                                  & $2.628\times 26.13^\circ \approx 6.9\,\text{deg}^2$        \\
				pixel scale                                    & $2.33''$/pix        \\
				Q.E.                                           & $60$\%              \\
				Gain                                           & 0.8                 \\
				Read noise                                     & 1.6 ADU             \\
				max frame rate                                 & 90\,Hz              \\ \hline
				\rule{0pt}{2.6ex}filter (F505W)                & 400--610\,nm tophat \\ 
				filter (F600W)                                 & 500--700\,nm tophat \\ \hline
 				\rule{0pt}{2.6ex}cutout size                   & $15\times 15$ pixels\\
				number of cutouts                              & $\mathcal{O}(1000)$ \\
				exposure time                                  & 39.5 ms             \\
				limiting mag. (39.5\,ms, $5\sigma$)            & 13.5                \\
				lim.~mag.~(100 stack) \rule[-0.9ex]{0pt}{0pt}  & 16                  \\
				lim.~mag.~(3 sec)                              & 18.5                \\
				lim.~mag.~(30 sec)                             & 19.5                \\ \hline
				\rule{0pt}{2.6ex}Optical track                 & 2066.5\,mm          \\
				Number of elements                             & 4 elements, 3 groups \\ \hline 
				\rule{0pt}{2.6ex}location                      & Mitzpe Ramon, Israel\\ 
				longitude                                      & $34^\circ45^{'}43.3^{''}$ E        \\ 
				latitude                                       & $30^\circ35^{'}48.4^{''}$ N        \\ 
				elevation                                      & 860\,m              
			\enddata
			\label{tab: properties}
						
		\end{deluxetable}

		Currently W-FAST is situated at the Wise Observatory near Mitzpe Ramon in the Negev desert in Israel, 
		and is expected to be relocated to a new site in Neot-Smadar, Israel, during 2021. 
		
		

		\subsection{Sensor}\label{sec: sensor}
		
		The detector system is a large format sCMOS sensor: 
		the \emph{Balor} camera from Andor\footnote{
			\url{andor.oxinst.com/products/scmos-camera-series/balor-scmos}}. 
		This camera has $4128\times 4104$\,pixels 
		with a pixel pitch of $12.2\times 12.2\,\mu$m, 
		for a total sensor size of $5\times 5$\,cm. 
		The camera can output images at a rate of 90\,Hz, 
		which is faster than required for the science goals of the experiment. 
		In our normal operation mode the data is read out at 10--25\,Hz, 
		with negligible dead-time. 
		
\begin{figure}
			
			\centering
			
			\pic[1]{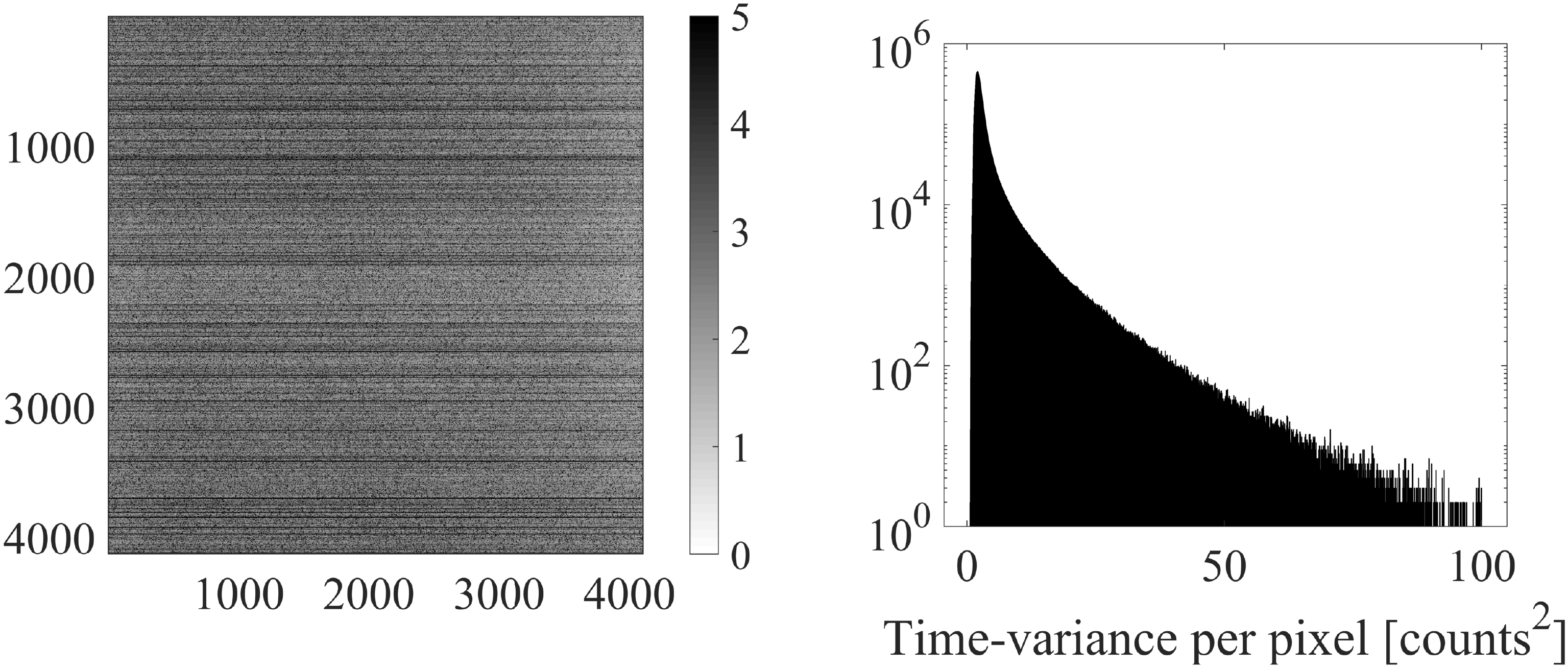}
			
			\caption{Individual pixel time-variance of the dark exposures 
				is calculated by taking the variance of each pixel
				along a batch of 100 dark frames, 
				and then averaging over all 203 batches. 
				Left: the map of the individual pixel time-variance. 
				There are clearly features along rows of the sensor. 
				Right: a histogram of the time-variance for different pixels, 
				excluding values below 0.5 and above 100 ADU$^2$. 
				While most of the pixels have a variance around 2.5 ADU$^2$, 
				a continuous tail of higher-variance pixels 
				makes it hard to define a cutoff value to mark as \textit{bad pixels}. 
			}
			\label{fig: dark time variance}
			
		\end{figure}

		The Quantum Efficiency (QE) of the sensor peaks at 60\% 
		in the range of 500--700\,nm, dropping to 40\% at $\approx 400$\,nm. 
		The dark current at $T\sim 0^\circ$\,C is between 0.1--0.25 ADU pixel$^{-1}$ s$^{-1}$, 
		which is below the mean sky brightness even during dark times, 
		which is around 10--25 ADU pixel$^{-1}$ s$^{-1}$. 
		For short exposures we are more concerned with the read noise, 
		which has a variance of around 2.5 ADU$^2$ for each frame. 
		This is true of the majority of pixels, 
		but many pixels show spatial correlations 
		and many pixels have a higher variance than the mean. 
		The average variance of each pixel in dark-frames is shown in 
		Figure~\ref{fig: dark time variance} (left)
		along with the distribution of pixel variances (right). 
		
		In addition, the spatial variance, calculated over a single image,
		is higher than the frame-to-frame variance (close to 4 ADU$^2$), 
		which is most likely due to the differences in the readout electronics of each pixel. 
		This additional variance accounts for some of the increased noise in the photometry
		relative to what we expect from the read noise values quoted above.
		The spatial variance affects the photometric stability
		when the speckle patterns of stars move around on different pixels 
		from one exposure to the next. 
		
		We choose to flag pixels with variance below 0.5 or above 50 counts$^2$, 
		as well as any pixels with mean higher than five times the standard deviation
		of the distribution of the pixel mean values. 
		We calculate the standard deviation using an iterative ``sigma clipping'' routine. 
		This flags a total of $\sim 26600$ pixels, only 0.15\% of the sensor.
		
		\subsection{Optical design}\label{sec: optical design}
		
\begin{figure*}
			
			\centering
			
			\pic[0.8]{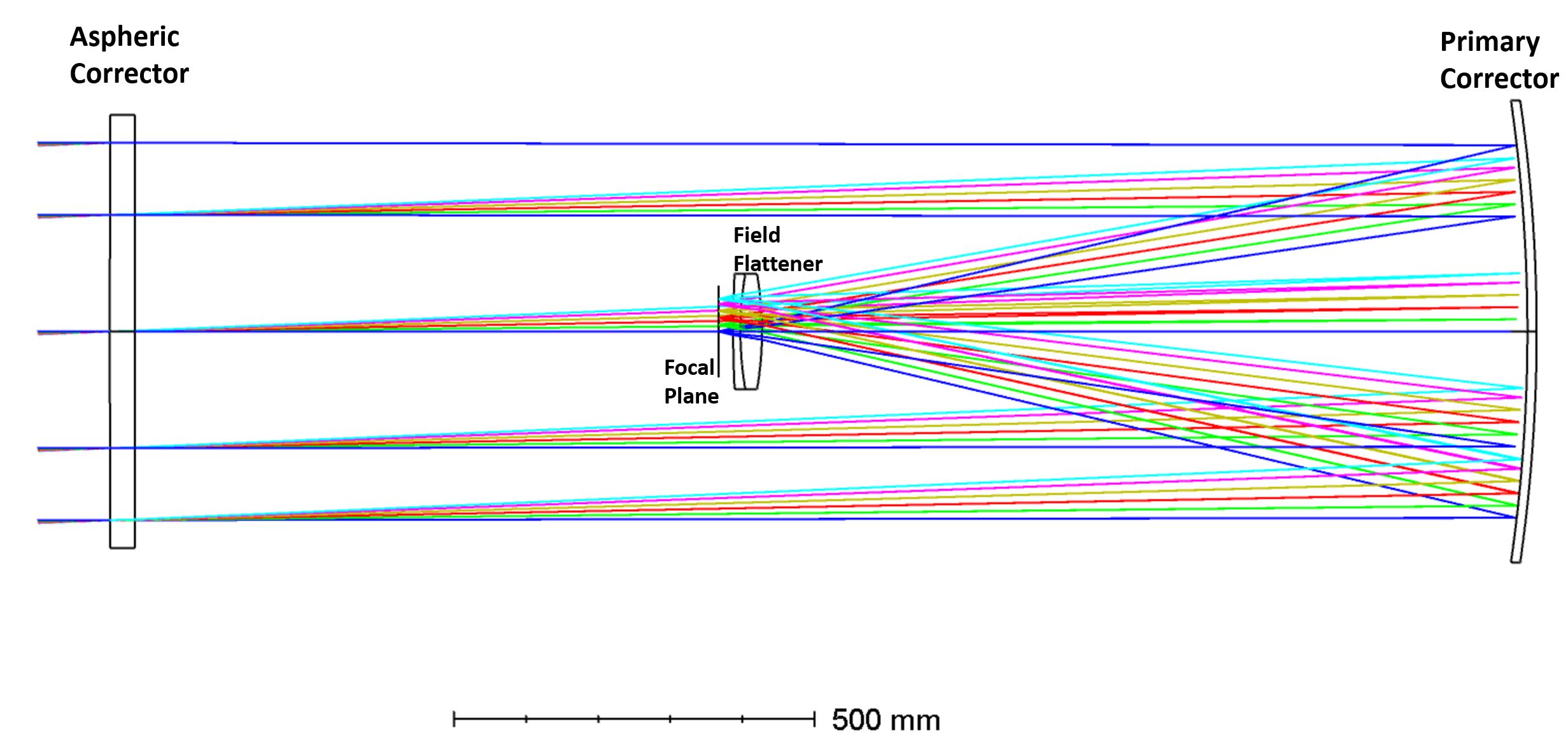}
			\caption{W-FAST telescope optical layout.}
			\label{fig: TelescopeLayout}
			
		\end{figure*}

		The W-FAST telescope is a Schmidt camera with an achromatic field flattener.
		In Figure~\ref{fig: TelescopeLayout} we show the optical layout of the system. 
		The first element in the telescope optical train is 
		an aspheric corrector plate ($6^{th}$ order even asphere). 
		Its substrate is a UV grade-A fused silica blank, 
		and the polished surfaces are coated with 
		a multi-layer dielectric anti-reflection coating. 
		The telescope primary mirror has a spherical surface figure. 
		We choose the Ohara Glass CCZ-HS low expansion ceramic 
		as the mirror substrate to minimize sensitivity to thermal fluctuations. 
		The mirror is coated with a layer of protected aluminum. 
		The last element before the detector unit is the field flattener, 
		a bonded BK7-SF2 achromatic lens. 
		A single filter is inserted before the field-flattener, 
		as discussed in Subsection~\ref{sec: filter}.
		All optical elements were fabricated by Arizona Optical Systems, LLC,\footnote{\url{https://arizonaoptics.com}} 
		except the filter which is manufactured by Asahi.\footnote{\url{https://www.asahi-spectra.com/}}
		
\begin{figure}
			
			\centering
			\pic[\picsize]{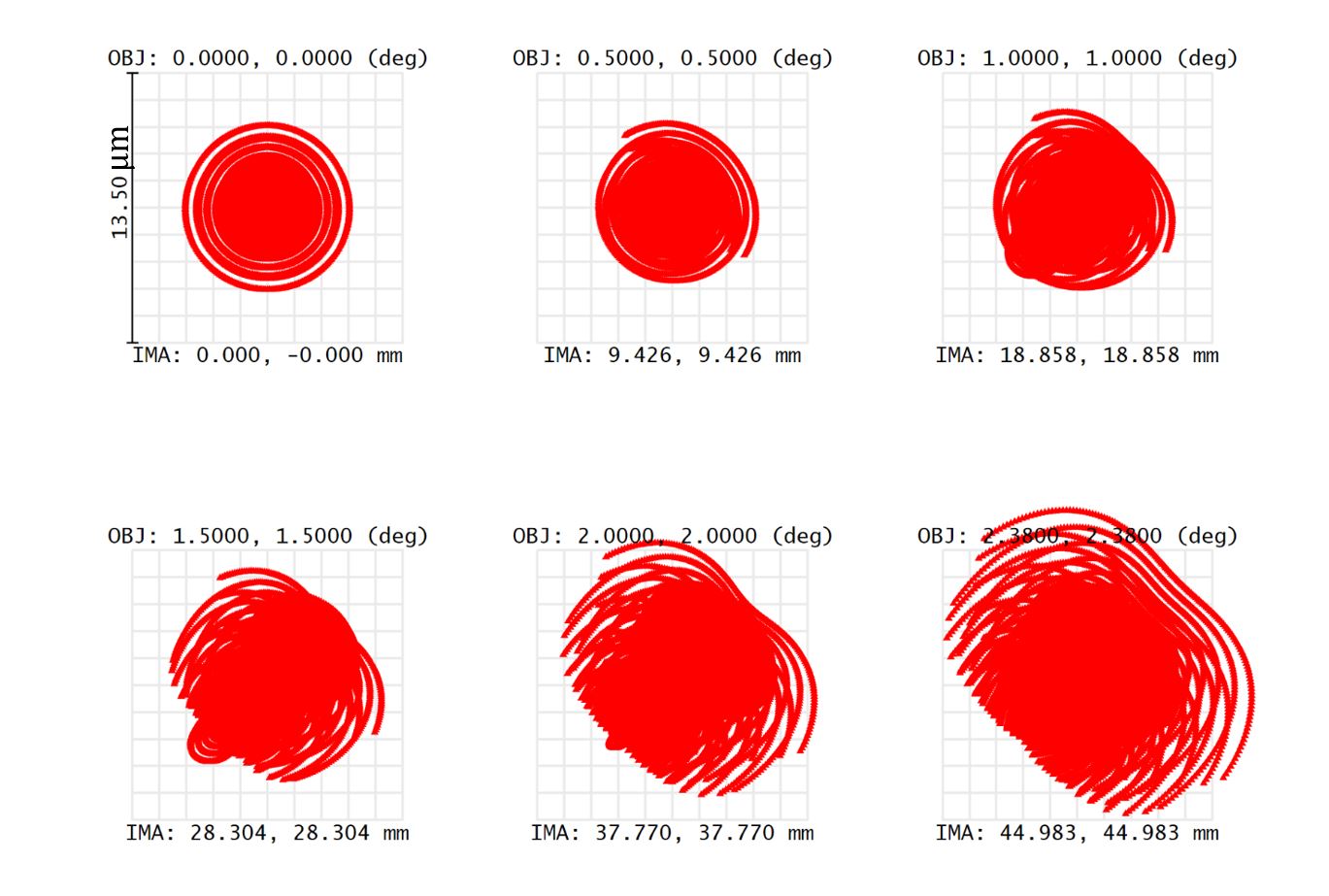}
			\caption{Telescope nominal (ray-tracing) image quality: spot diagrams in Sloan r-band.
			         The squares in the images correspond to 1.35\,$\mu$m, or $0.2578''$.}
			\label{fig: TelescopeSpots}
			
		\end{figure}

\begin{figure}
			
			\centering
			
			\pic[\picsize]{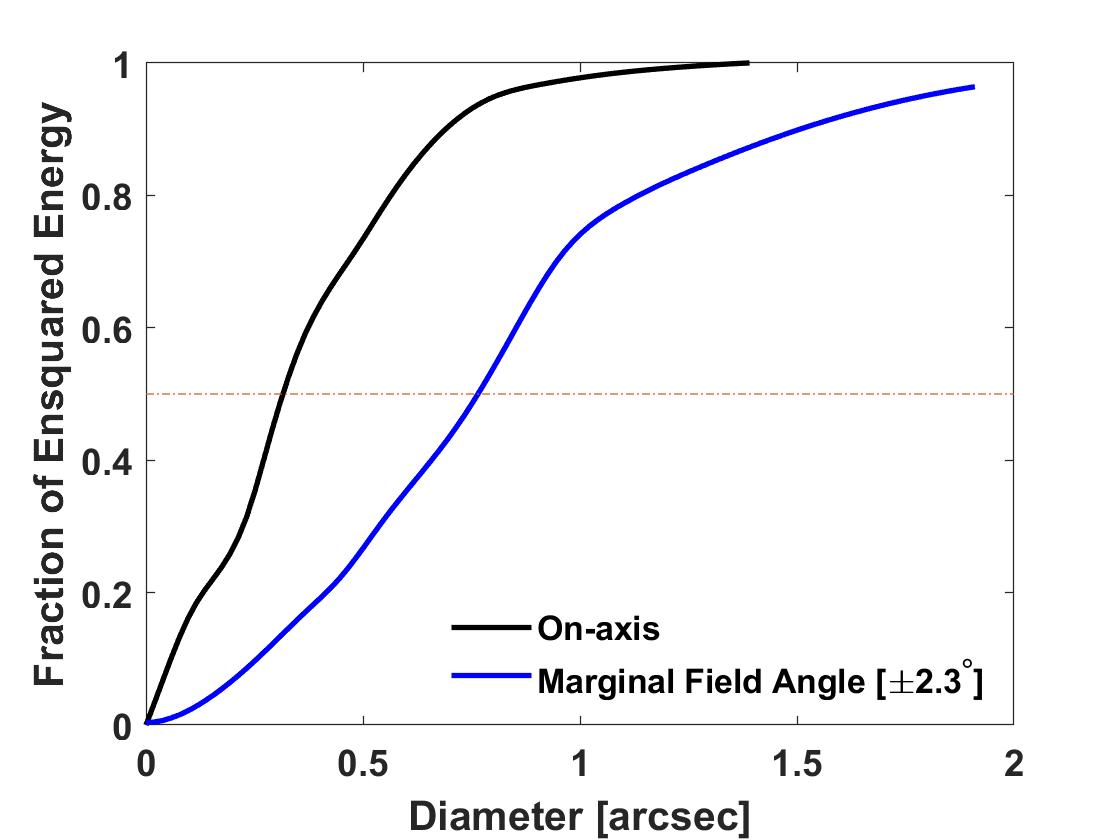}
			
			\caption{Telescope nominal (ray-tracing) image quality: 
				fraction energy ensquared in a range of spot diameters. 
				Calculated in Sloan r-band for on-axis and marginal field angle.}
			\label{fig: EnsquaredEnergy}
			
		\end{figure}

		Figure~\ref{fig: TelescopeSpots} presents the theoretical spot diagrams,
		while Figure~\ref{fig: EnsquaredEnergy} shows the full width at half maximum (approximated via
		50\% encircled energy), as a function of position in the focal plane.
		Ignoring manufacturing and assembly inaccuracies as well as
		the seeing conditions and detector effects, 
		the spot sizes are under-sampled when used 
		with the a 12\,$\mu$m pixel camera.  
		The leading Seidel aberration is distortion (barrel), 
		as is expected in such a large FoV system. 
		The ensquared energy for on-axis and marginal-field angles 
		show that the image quality will be governed by 
		seeing conditions in all realistic scenarios.
		These estimates for the image quality 
		are based on the redder, 500--700\,nm filter. 
		With the bluer, 400--610\,nm filter the image quality 
		is worse, giving spot sizes comparable to 
		the pixel size or the atmospheric seeing. 
		
		\begin{figure*}
			\centering

			\pic[0.25]{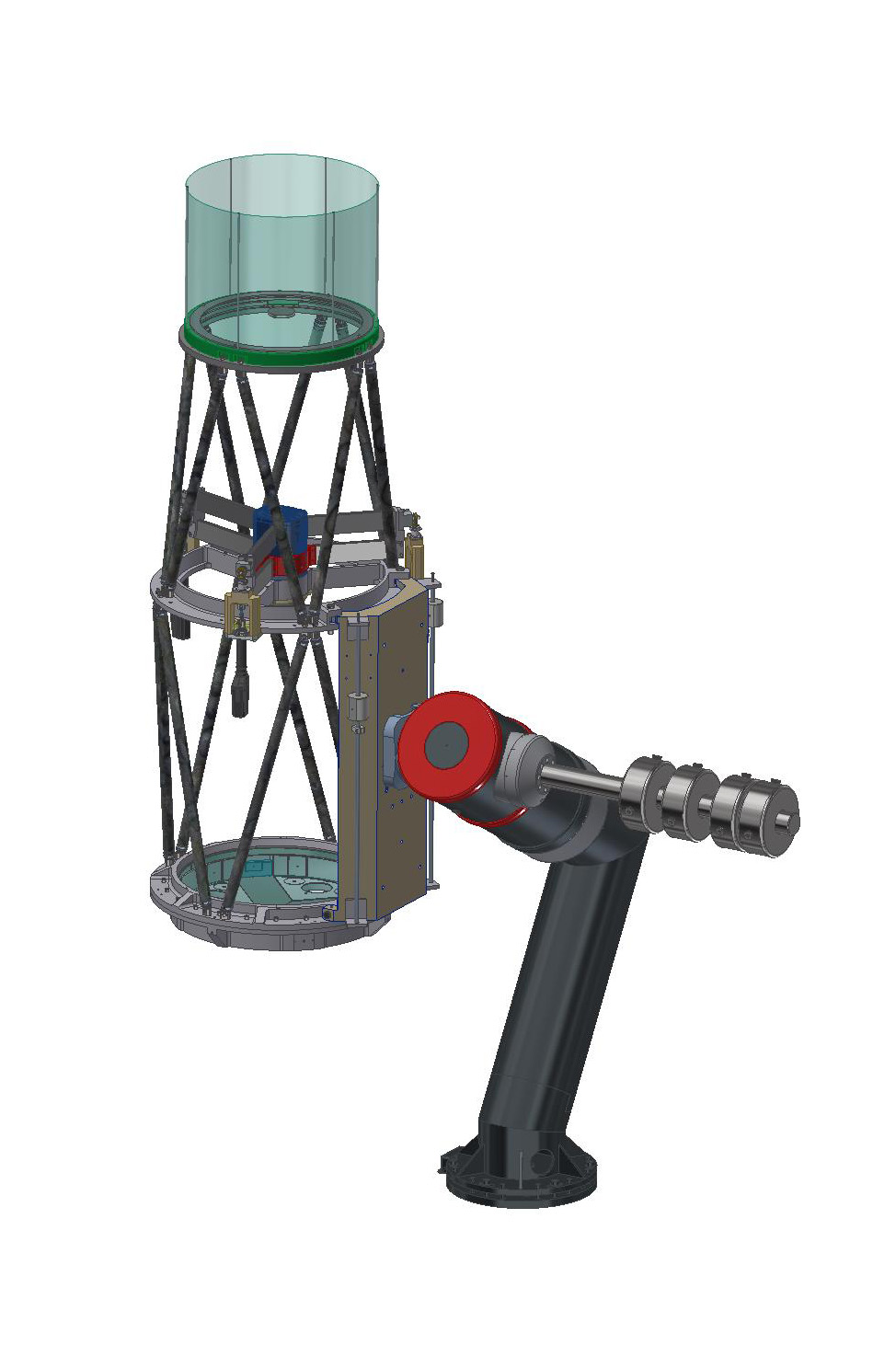}
			\hspace{1cm}
			\pic[0.25]{WFAST_telescope}

			\caption{The W-FAST optical tube assembly. 
				Left: CAD image of the mechanical design. 
				Right: the telescope after installation at the observatory. }
			\label{fig: telescope assembly}
			
		\end{figure*}

		\subsection{Filter}\label{sec: filter}
		
\begin{figure}
			\centering
			\pic[0.8]{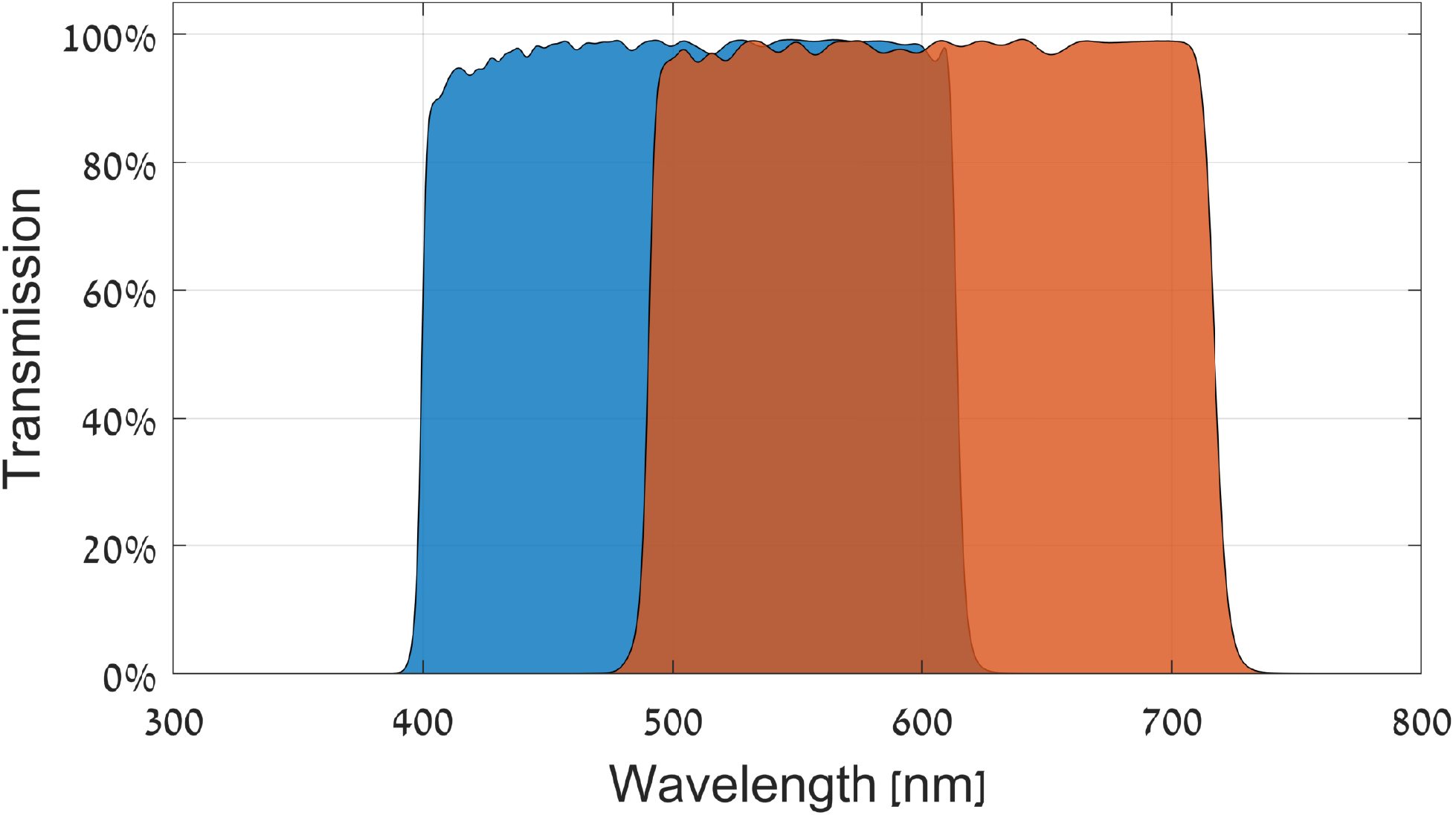}
			\caption{Transmission curves for the F505W and F600W filters. See Table~\ref{tab: filter transmission}.}
			\label{fig: filter transmission}
		\end{figure}

		The telescope is equipped with a single filter at any given time. 
		There are two available filters that can be manually exchanged. 
		The F505W from Asahi, 
		has a top-hat transmission curve between 400 to 610\,nm. 
		This filter was chosen to maximize throughput at the blue side of the
		visible spectrum (where the Fresnel scale is smaller). 
		and given in Table~\ref{tab: filter transmission}.
		A second filter, F600W, covers the range 500--700\,nm. 
		The measured filter transmissions are 
		shown in Figure~\ref{fig: filter transmission}.
		
		\begin{deluxetable}{ccc}
			
			\tabletypesize{\footnotesize}
			\tablewidth{0pt}
			\tablecaption{Transmission curve for the two W-FAST filters.}
			\tablecomments{Filter transmission for F505W (the bluer filter) and the F600W (the red filter). 
				The complete Table~\ref{tab: filter transmission} is published in its entirety in a machine-readable format.
				}
			
			\tablehead{
				\colhead{Wavelength [nm]} & \colhead{F505W [\%]} & \colhead{F600W [\%]} 
			}
			
			\startdata 


			1100.0 & 0.017 & 0.002 \\ 
			1099.5 & 0.017 & 0.001 \\ 
			1099.0 & 0.017 & 0.002 \\ 
			1098.5 & 0.017 & 0.002 \\ 
			1098.0 & 0.017 & 0.002 \\ 
			...    & ...   & ... \\ 
			\enddata
			
			\label{tab: filter transmission}
			
			
		\end{deluxetable}
		
		\subsection{Seeing}\label{sec: seeing}
		
		The median seeing in the Neot Smadar site is $\sim 1.5''$, 
		which is slightly smaller than the pixel scale of $2.29''$. 
		At the Mitzpe Ramon site the seeing is somewhat worse, $\sim 2.5''$. 
		The optical quality is reduced when using the bluer, F505W filter, 
		and the spot size becomes comparable to the atmospheric seeing. 
		The combined pixel size, atmospheric and instrumental seeing 
		gives a total of 3.5--$5''$ FWHM in normal observing conditions. 
		
		\subsection{Optical Tube Assembly}\label{sec: optical tube assembly}
		
		The Optical Tube Assembly (OTA) has been designed and manufactured by the Weizmann Institute. 
		It is based on a three-level truss support, 
		where each level is held up by eight carbon fiber tubes. 
		The primary mirror is held inside a blackened aluminum housing on the bottom level 
		that is kept far enough from the mirror to allow thermal contraction of the aluminum, 
		using 2\,mm high-stiffness silicon pads.
		The silicon pads are connected to an Invar cross 
		that provides a non-expanding interface to the aluminum housing. 
		The pads are connected to the Invar using 
		aluminum holdings that can be shifted in the $xy$ plane using micrometric screws 
		for controlling the de-center of the primary mirror. 
		The top level of the OTA  holds the corrector lens in an Invar ring with silicon pads. 
		The middle level, in the prime focus of the telescope, 
		has a three legged spider holding the field flattener and filter underneath it. 
		On top of that is a second spider made with very rigid silicon-carbide arms, 
		that holds the camera just a few millimeters above the field flattener. 
		The second spider is held up by three Physik Instrumente MA-35 linear actuators\footnote{
		\url{https://www.physikinstrumente.com}.} 
		that control the camera focus and tip-tilt corrections at the micrometer level. 
		The width of each spider arm is less than 1\,cm.
		The camera and field flattener cause a circular obscuration with a diameter of 19\,cm.
		The total mass of the OTA, including optics, electronics, 
		and the plate connecting the telescope to the mount
		is approximately 280\,kg. 
		
		\subsection{Mount}
		
		The mount used is the Astro Systeme Austria\footnote{\url{https://www.astrosysteme.com/}} (ASA) DDM160, 
		a German-Equatorial mount working close to the maximum weight limit of 300\,kg on the telescope side. 
		There are additional sliding weights on the side of the OTA used to balance the telescope
		after adding or removing equipment. 
		The mount can rotate more than 45 degrees beyond the meridian. 
		The minimal (software) altitude for observations is $25^\circ$ during normal operations 
		limiting the minimal declination from our observatory to $\delta\leq-35^\circ$.
		
		In addition to the mount control software, we added Arduino chips with accelerometers (ADXL345) 
		to provide rough estimate of the pointing of the telescope vs.~the Earth's gravity vector. 
		This prevents crashes of the front of the telescope in case of software failure. 
		
		\subsection{Observatory}\label{sec: observatory}
		
\begin{figure}
			\centering
			\pic[0.5]{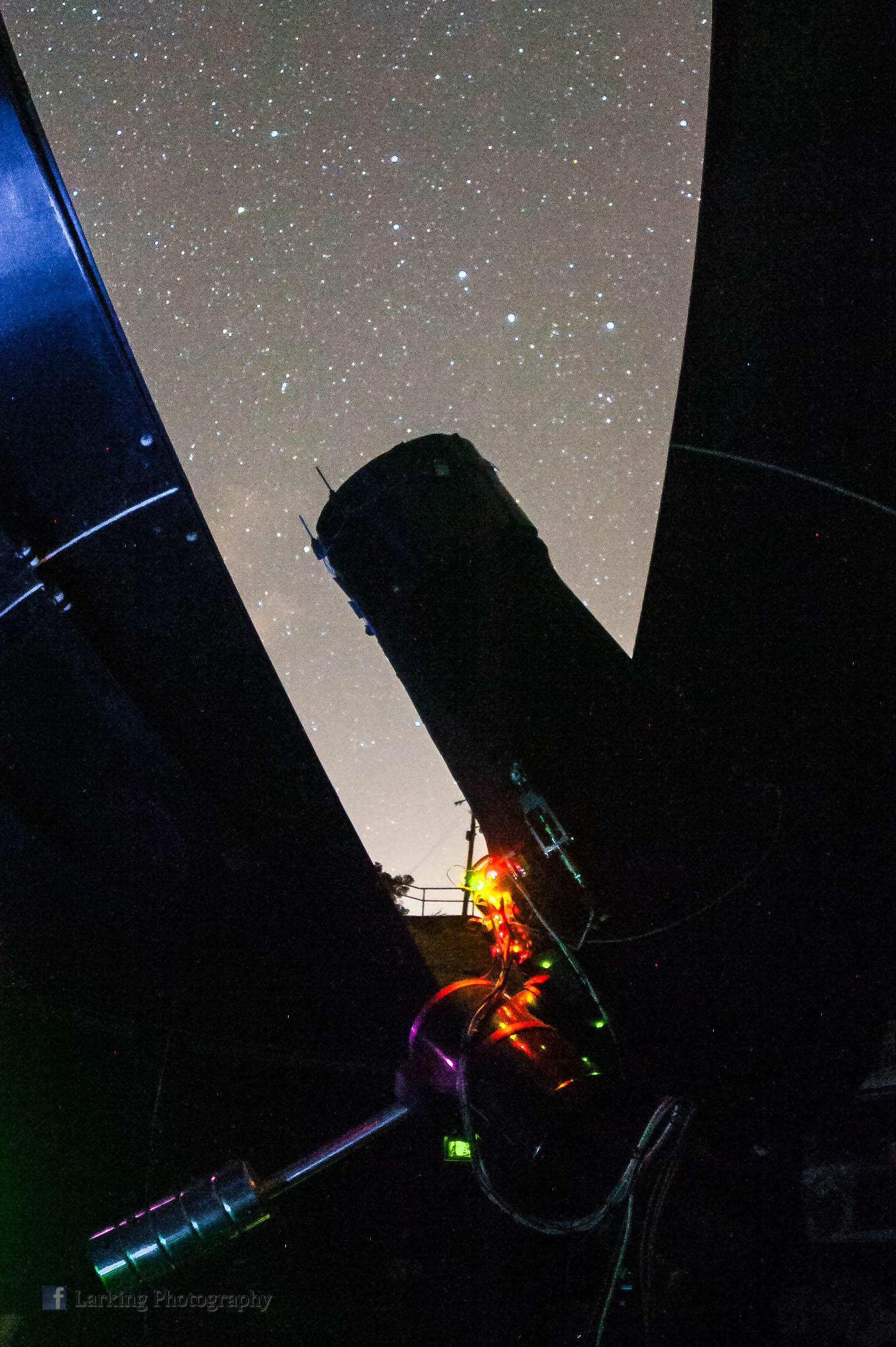}
			\caption{An image from the dome DSLR camera. 
				This camera is used to make sure the telescope and dome are positioned properly, 
				and to verify the sky conditions.}
			\label{fig: dome slr}
		\end{figure}

		The dome is a 18ft (5.5m) diameter clam-shell design fiberglass dome from AstroHaven\footnote{\url{https://astrohaven.com/}}. 
		Outside the dome is a weather sensor: a Boltwood Cloud Sensor II weather station\footnote{\url{https://diffractionlimited.com/}}.
		Inside the dome we have added a humidity, temperature and pressure sensors, 
		and a set of cameras to monitor the telescope operations. 
		A DSLR camera (Sony Alpha 7RII) 
		equipped with a wide angle lens (12mm fish-eye) 
		allows checking the telescope and dome positions, as well as verifying 
		by eye the sky brightness and the visibility of stars. 
		An example image from the dome DSLR is shown in Figure~\ref{fig: dome slr}. 
		In addition, 
		we monitor the weather sensors from other observatories on site, 
		and use multiple measurements to decide if the dome should be closed because of bad weather. 
		
	\section{Observatory control system}\label{sec: observatory control system}
		
		The observatory is controlled by two computers running Windows.
		Control of most components is handled through matlab, 
		either through ASCOM drivers loaded as objects into the matlab environment, 
		through serial or TCP/IP ports, or using C++/mex functions to interface with 
		device SDKs. 
		
		The first computer manages the control of the dome, mount, 
		and weather sensors, scheduler and autonomous observatory control. 
		This control computer communicates with a dedicated camera computer	via TCP/IP ports, 
		mainly for camera control and auto-guiding. 

		The second computer is dedicated to controlling the data acquisition 
		and the focus system. 
		This computer has 32 cores on a single CPU chip, 
		256 GB of RAM and a high read-/write-speed SSD drive with 2 TB of storage, 
		that is used as a buffer for each night's acquisition. 
		This data is compressed and backed up to two 14 TB drives, 
		one of which is detachable so it can be switched out every few weeks 
		to transfer the data back to the Weizmann Institute servers. 
		
		\subsection{Scheduling and robotic operations}\label{sec: scheduling}
		
		Targets are selected using a custom software from a list of pre-defined fields. 
		At specified intervals the scheduler program checks the target bank
		and makes a decision if to continue the observations or to switch targets. 
		The main parameters used for decision making are the user-defined priority, 
		the airmass, and any other observational constraints, e.g., moon. 
		The scheduler also takes into account the weather conditions, 
		and will not observe, e.g., Westerly targets when Western winds are too strong. 
		Switching fields usually takes a few minutes, 
		with most of the time spent on focusing. 
		These constraints usually mean that fields are observed 
		for 1--2 hours before reaching the meridian, 
		and 1--2 hours after, 
		staying most of the time, by design, 
		at the lowest airmass. 
		A simulator program allows the user to 
		see the targets for upcoming nights, 
		and make adjustments to the priorities or observational constraints. 
		
%
%
%


		
		
	\section{Data acquisition and processing}\label{sec: data acquisition and processing}
		
		\subsection{Data acquisition}\label{sec: data acquisition}
		
		The images are acquired using a custom interface between matlab and the Andor SDK. 
		The camera is set to operate continuously on a separate thread, 
		filling batches of 100 images to buffers that are then read out
		by the main thread of matlab for further processing. 
		Because of the high data rate ($\approx 6$\,Gb s$^{-1}$), 
		only a small part of the full data set can be processed. 
		The data reduction pipeline produces two primary data products: 
		cutouts and stack images. 
		An additional mode for writing bursts of full-frame images 
		is also available, although local hard drive space limits this
		mode to less than an hour of data. 
		
		The cutouts are square with a few pixels on a side around the position of each star. 
		The positions are measured in the beginning of each run using a stack of a 100 images, 
		then the positions are updated at the beginning of each batch of 100 images 
		to compensate for telescope drift
		based on the centroids measured from all the cutouts. 
		At the beginning of each run the star positions
		are used to obtain an astrometric solution 
		using code from \cite{astrometry_code_Ofek_2019}
		and the stars are cross matched using code from 
		\cite{catsHTM_Soumagnac_Ofek_2018}, 
		based on the GAIA DR2 catalog
		\citep{GAIA_data_release_two_2018}. 
		
\begin{figure}
			
			\centering
			\pic[\picsize]{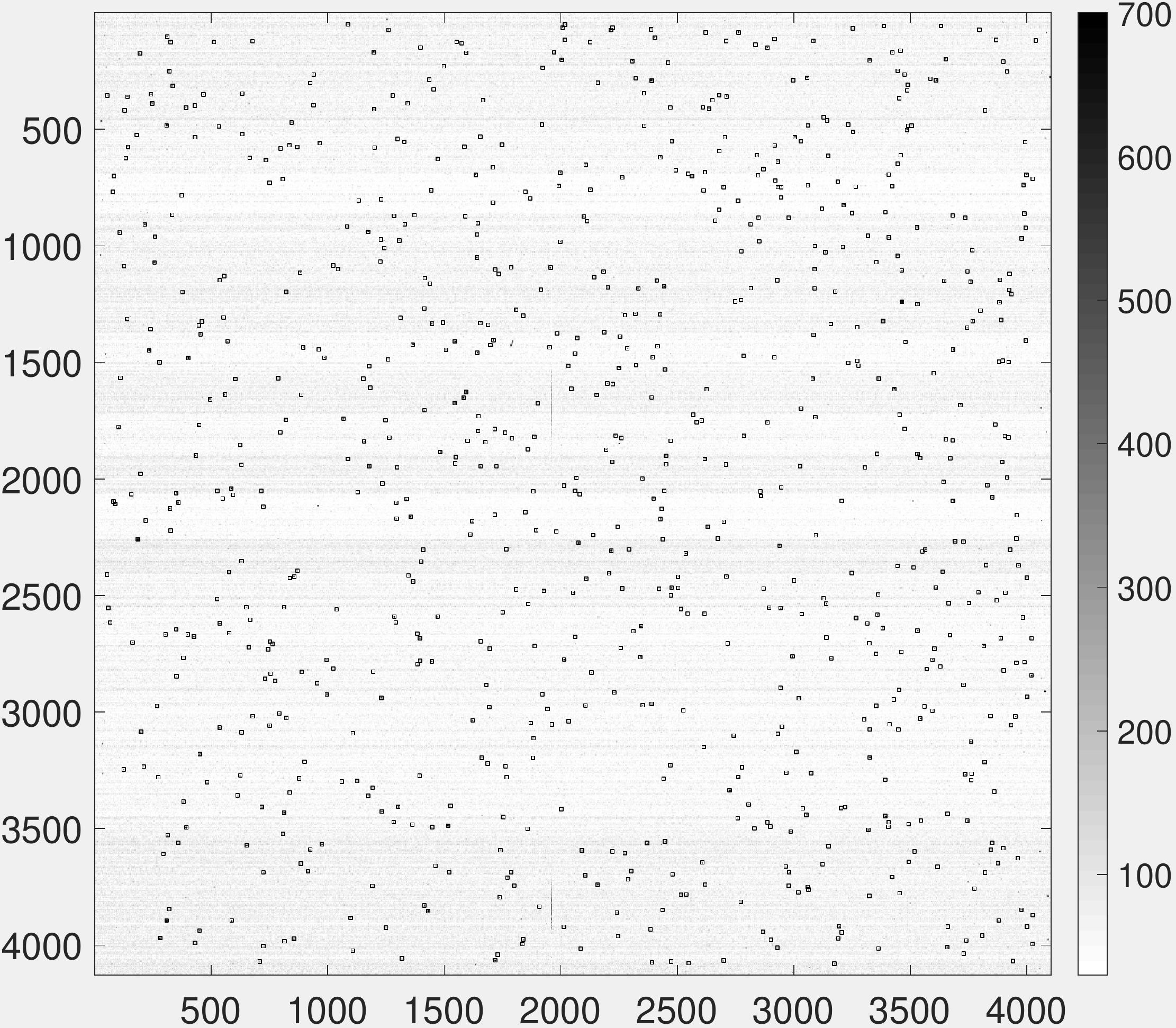}
			\caption{Example W-FAST field with cutouts 
				marked around the positions of 2000 stars. 
				The vertical line at $x=1960$ is a bad column in the CMOS. 
				The horizontal stripes are due to correlations in the read noise. 
				The color-scale has been inverted to allow easier viewing. 
			}
			\label{fig: balor cutouts full}
		\end{figure}

\begin{figure}
		
		\centering
		\pic[\picsize]{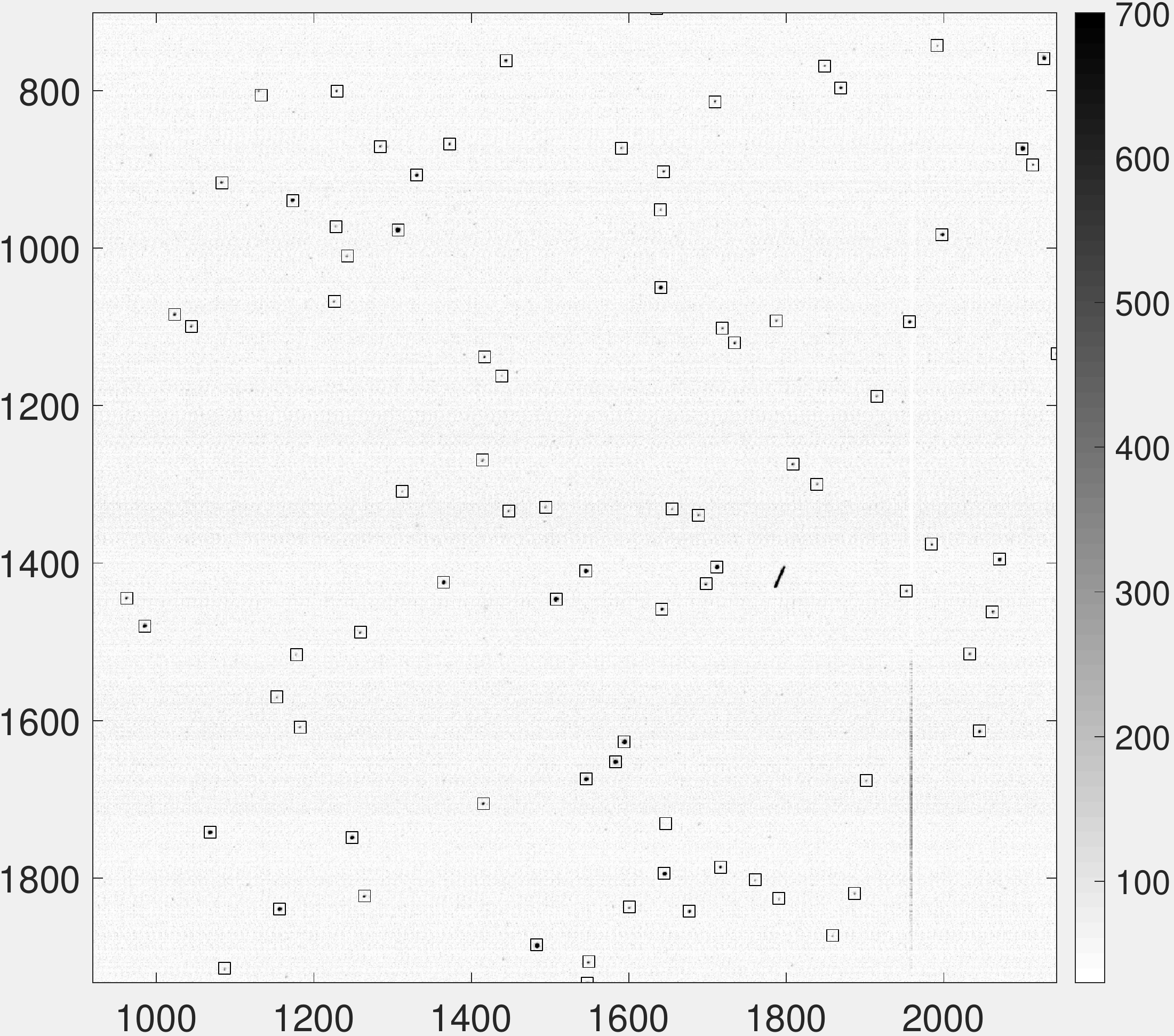}
		\caption{The same as Figure~\ref{fig: balor cutouts full} 
			but zoomed in on a small region of the sensor. 
			The stars are visible inside the $15\times 15$ pixel cutouts, 
			as well as a satellite streak. 
		}
		\label{fig: balor cutouts closeup}
	\end{figure}

		Cutouts are extracted to a separate 4D matrix,
		which is dark- and flat-calibrated.
		Rough photometry is done on the calibrated cutouts during the run, 
		simply summing all the pixel values in each cutout, 
		while more precise photometry is done offline during the day. 
		An example image with cutout boundaries is shown in
		Figure~\ref{fig: balor cutouts full}. 
		The black squares are cutouts around the 2000 stars in this field. 
		A closeup of the same field is shown in Figure~\ref{fig: balor cutouts closeup}, 
		showing the stars inside the cutouts as well as a satellite streak. 
		
		The coadded images (stacks) images are used to do science on dimmer targets 
		at lower cadence. 
		The stacked are produced by summing 
		the full-frame images directly into a single precision floating point array, 
		that are dark- and flat- calibrated after the summation. 

		Once the data is reduced to cutouts and stacks, 
		its volume is reduced to about 5\% of the initial size
		and it can be saved to disk 
		to be handled offline by more complicated code. 
		
		The image calibration includes dark, flat
		and bad-pixel flagging. 
		A separate calibration class is used to add together 
		hundreds of full-frame images taken when recording 
		darks or flats and used to produce four maps:
		(a) dark mean values (dark map), 
		(b) dark variance map, 
		(c) flat mean values, 
		(d) bad pixel map. 
		The bad pixels are defined as pixels that have a mean value
		larger or smaller than the average bias level by more than 5$\sigma$
		using an iterative ``sigma clipping" algorithm with five iterations. 
		In addition, pixels with variance smaller than 0.5 or larger than 50 counts$^2$ 
		are considered bad pixels. 
		The calibration object keeps a set of cutouts with the same size
		as the image cutouts of each of these maps, 
		which is updated upon each batch of input images to the new positions of the stars, 
		and used when calibrating the cutout images. 
		The dark and flat cutouts are then applied in the standard way to the image cutouts. 
		
		The stack image is bias subtracted using the dark mean values multiplied 
		by the number of images used to make the stack, then flat-fielded. 
		Finally, both cutouts and stack images have bad pixels replaced by NaN values.
		
		Cutouts are also produced from the stack images, 
		and used to measure telescope drift every few seconds for auto-guiding. 
		Aperture photometry is done on the stack-cutouts to check that the stars are still visible. 
		If a large fraction of stars lose a large fraction of light compared to the previous ten measurements, 
		the positions of cutouts is rechecked against the full frame stack, 
		using a custom ``quick align" routine. 
		The quick align sums the stack image columns and rows, 
		producing two vectors with peaks on the row or column containing bright stars. 
		These vectors are cross correlated 
		with vectors produced in the beginning of the run, using FFT. 
		The peak of the correlation is used to quickly recover the shift in $x$ and $y$ 
		of the new image relative to the reference stack. 
		This method allows the cutouts to be shifted quickly into place 
		even after large drifts in telescope position. 
		A very similar method is described in \cite{donuts_autoguiding_McCormac_2013}. 
		If the stars are not visible after the quick align, the batch is aborted (e.g., when clouds roll in). 
		If more than four batches are aborted the run is aborted. 
		
		W-FAST can be operated in one of two principle modes: 
		(1) The high-cadence mode described above, producing cutouts around bright stars 
		and stacked full frame images at lower cadence;
		(2) A low cadence mode where full frame images are saved every few seconds (3 seconds is the default setting), 
		to capture fainter objects. 
		Longer exposures can be built up from 3-second images. 
		This allows image alignment and cosmic ray removal before adding the images into deeper co-adds. 
		
		In the high-cadence mode, the stack images have an integrated exposure of four seconds, 
		and are about 10 times deeper than the single images. 
		Since the read noise RMS is $\approx 1.6$\,e$^-$ pixel$^{-1}$ and 
		the sky background at 25\,Hz is $\approx 0.4$\,e$^-$ pixel$^{-1}$, 
		the effect of stacking the short exposures increases the noise in the images by a factor of $\sim 4$. 
		This and other systematics like atmospheric and tracking jitter 
		leads to a loss of about two magnitudes when looking at stacks of 25\,Hz images, 
		compared to the native three-second exposures. 
		We measure the background on each individual cutout, 
		so we can correct for gradients in the background illumination, 
		e.g., when observing near the moon. 
		
		\subsection{Offline analysis}\label{sec: offline analysis}
		
		The cutouts are saved without calibration, 
		so the first step after loading each batch from disk is to perform dark and flat calibration. 
		Aperture photometry with adjusted aperture centers 
		is performed on each star in each frame using
		a custom-built, multi-threaded mex/C++ photometry code.
		We use aperture photometry which is more stable than PSF photometry for such short exposures. 
		We use an aperture radius of $7''$ to minimize instances of momentary loss of star light, 
		that could occur when the PSF is shifted or expanded. 
		
		The photometry pipeline calculates flux and the star's centroids 
		($1^\text{st}$ moment) of each star in each frame, 
		using the pixel values weighted by a centered Gaussian. 
		Then the aperture position is updated based on the centroid positions in each 
		cutout and the photometry is repeated. 
		The flux-weighted mean offsets of all stars inside their cutouts is used
		to position the final aperture used when calculating all photometric products. 
		This {\it forced photometry} is used since most of the stars' motion inside the cutout
		is due to tracking errors, so it is uniform across the image. 
		Using forced photometry makes sure that even faint stars have a properly positioned aperture. 
		An annulus with inner and outer radii of $17''$ and $23''$,
		concentric with the aperture for each star,
		is used for background estimation. 
		Finally, the number of bad pixels in each aperture is saved for each star and each image. 
		
		The flux, background, centroids, width ($2^\text{nd}$ moment) and number of bad pixels, 
		calculated in the repositioned aperture, 
		are saved as auxiliary data along with the fluxes.

\begin{figure}
			
			\centering
			
			\pic[1]{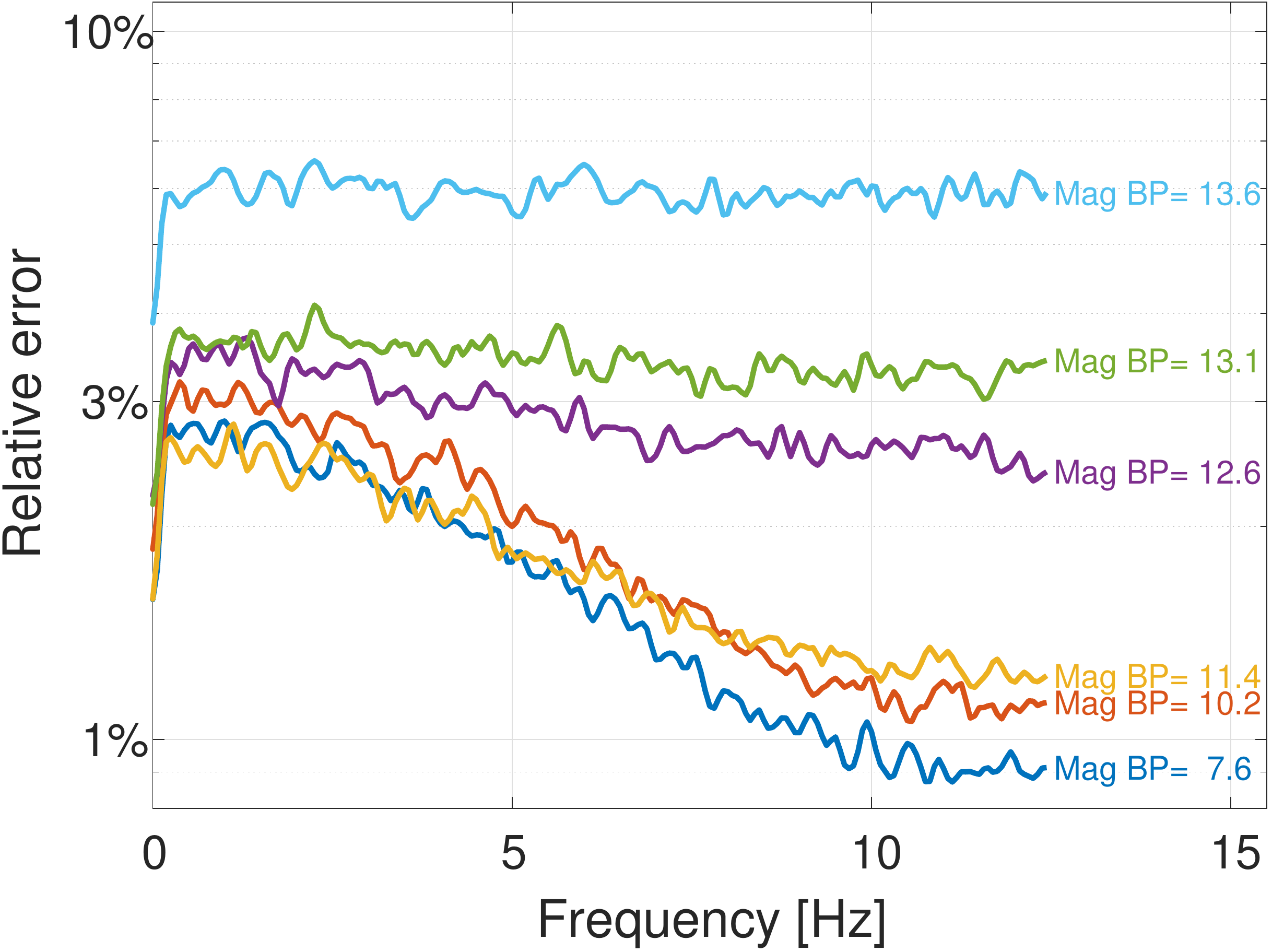}
			\caption{Example estimate for the Power Spectral Density (PSD). 
				The PSD is estimated using Welch's method on $2\times 10^4$ 
				flux samples for each star. 
				We show the square root of the power spectrum (integrated over frequency bins), 
				normalized to the average flux of each star. 
				We show this relative error per frequency for a few sample stars. 
				The magnitude of each star is mentioned in the plot, 
				showing the differences between bright and faint stars. 
				The bright stars have a clear red-noise component, 
				that can be reduced by applying a PSD correction. 
			}
			\label{fig: psd example}

		\end{figure}

		The resulting lightcurves show correlated noise on all time-scales, 
		with stronger noise at longer time-scales (red noise). 
		This reduces the effectiveness of our event finding algorithms, 
		as explained by \cite{LIGO_detections_PSD_correction_Zackay_2019}. 
		To mitigate the effects of red noise, 
		we calculate the Power Spectral Density (PSD) 
		using Welch's method \citep{power_spectral_density_Welch_1967}, 
		over $10^4$ flux measurements for each star. 
		The window size is set to 200 samples, with 50\% overlap between windows. 
		A few examples for the shape of the PSD are shown in Figure~\ref{fig: psd example}. 
		The noise of faint stars is dominated by uncorrelated noise, 
		presumably from read noise, 
		while bright stars have more structure in their PSD, 
		suggesting scintillation and telescope drift over inhomogeneous pixels
		dominate the noise budget for these targets.

		\subsection{Limiting magnitude}\label{sec: limiting magnitude}

		We conducted test observations during dark nights
		on 2021 Arpil 20, choosing a field
		at (J2000) $\alpha= 14^h59^m57.3^s$ and $\delta= +32^\circ43^{'}16.0^{''}$, 
		with an airmass of $\approx 1$. 
		We observed for a few minutes in several modes. 
		To estimate the limiting magnitude we plot the detection S/N
		(the ratio of the measured flux to the background in that region of the image)
		vs.~the Gaia DR2 magnitude for each star. 
		Then we fit the logarithm of the S/N to the magnitudes using a linear function, 
		iteratively removing outliers further than 2.5 times the scatter around the fit. 
		Stars without matches were also excluded from the fit. 
		Then we interpolated the point on the linear function where S/N$=5$
		and record the magnitude value for that point as the limiting magnitude. 
		
\begin{figure}
			
			\centering
			
			\pic[1]{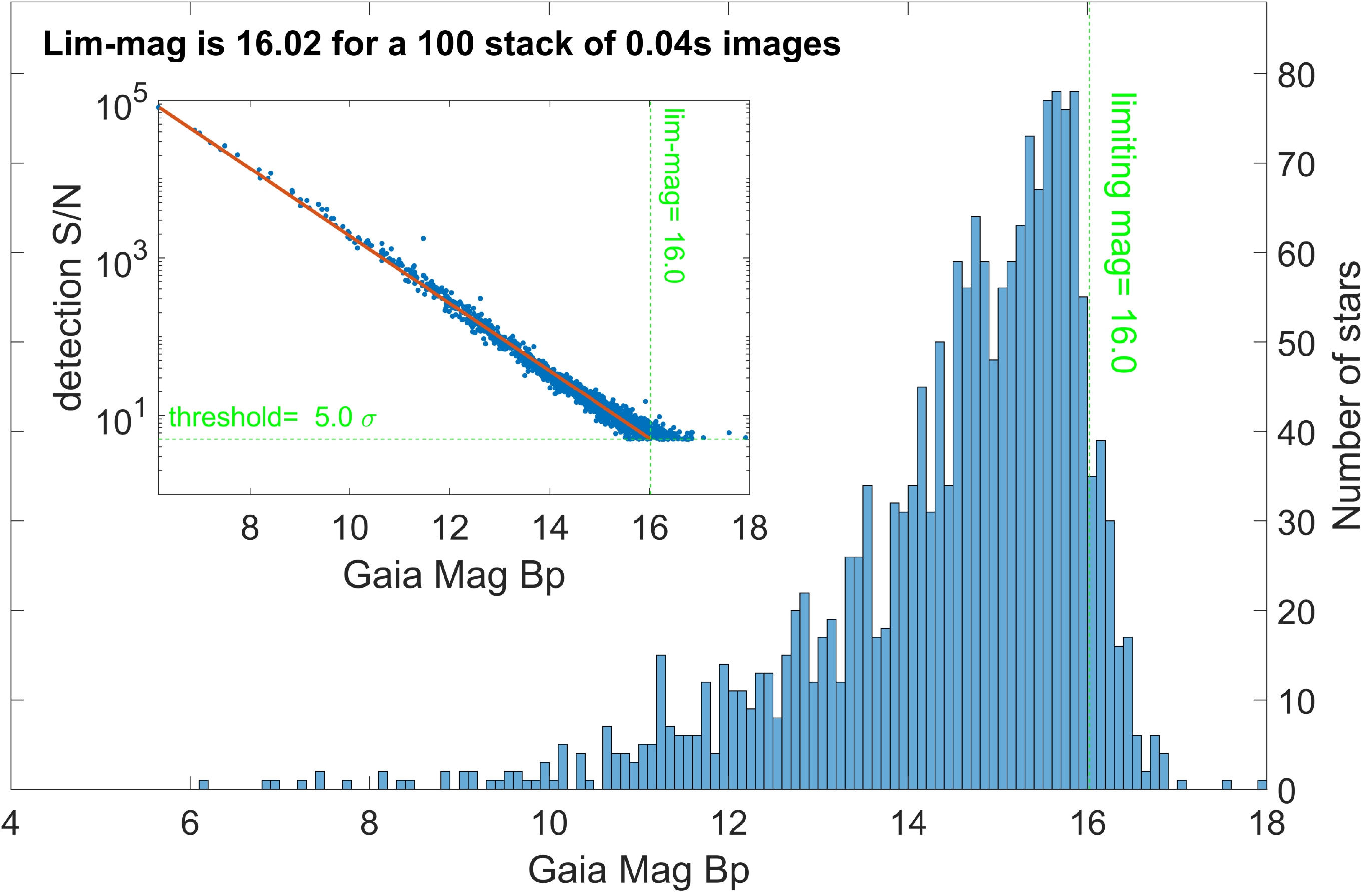}
			\caption{Histogram of matched stars with their GAIA DR2 magnitudes. 
				The inset shows a fit to the matched stars' $BP$ magnitude vs.~detection signal to noise (S/N). 
				We fit a 2nd order polynomial and then interpolate the fit to S/N$=5$, 
				giving the limiting magnitude of 16.0\,mag. 
			}
			\label{fig: lim_mag_25Hz}
		\end{figure}

\begin{figure}
			
			\centering
			
			\pic[1]{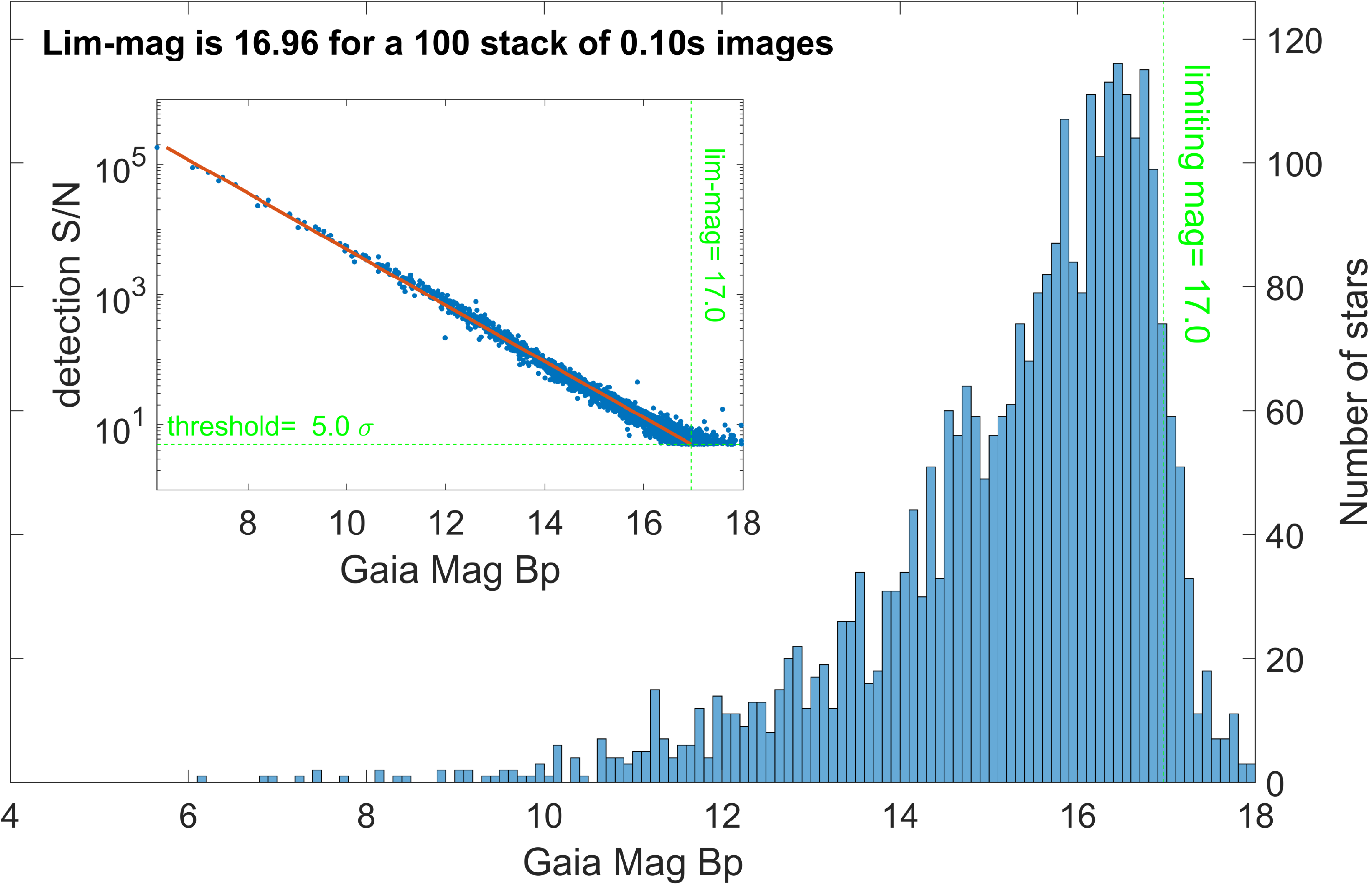}
			\caption{Same as Figure~\ref{fig: lim_mag_25Hz} only for a 0.1\,s exposures (taken at 10\,Hz). 
				The limiting magnitude from the fit is 17.0\,mag.}			
			\label{fig: lim_mag_10Hz}
			
		\end{figure}

\begin{figure}
			
			\centering
			
			\pic[1]{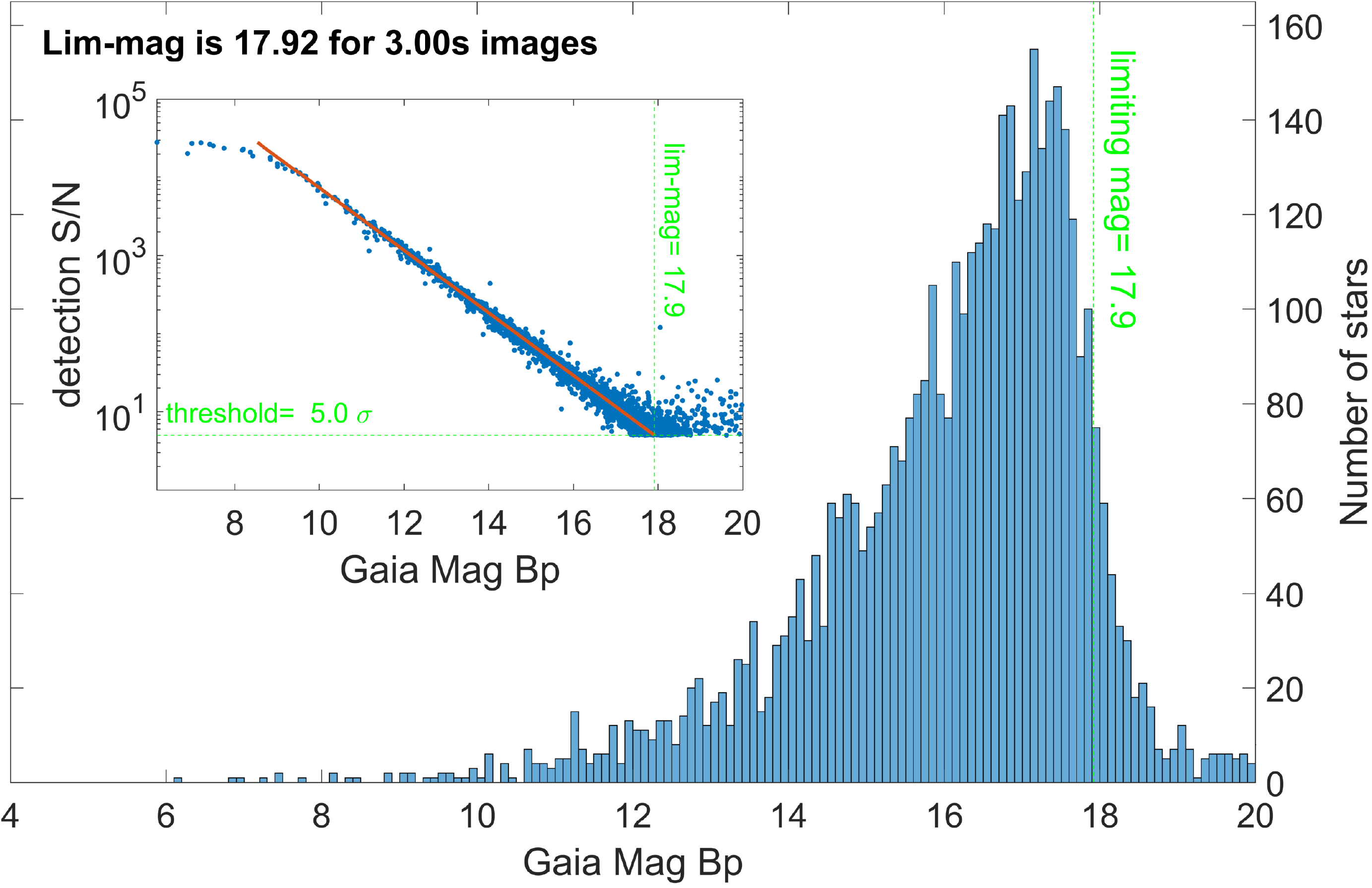}
			\caption{Same as Figure~\ref{fig: lim_mag_25Hz} only for a single, 3 second exposure.
				The limiting magnitude from the fit is 17.9\,mag.}
			\label{fig: lim_mag_3s}
		\end{figure}

		The limiting magnitude values are summarized in Table~\ref{tab: limiting mag}
		and the distributions used to find the limiting magnitudes for some of these
		cases are shown in Figures~\ref{fig: lim_mag_25Hz}-\ref{fig: lim_mag_30s}). 
				
		\begin{deluxetable}{lc}
			
			\tabletypesize{\footnotesize}
			\tablewidth{0pt}
			\tablecaption{Limiting magnitude at various exposure times}
			
			\tablehead{
				\colhead{Exposure} & \colhead{$5\sigma$ limiting magnitude} 
			}
			
			\startdata 
				40\,ms (single) & 14.5 \\
				40\,ms (100 stack) & 16.0 \\
				100\,ms (single) & 15.5 \\
				100\,ms (100 stack) & 17.0 \\
				1\,s (single) & 17.5 \\
				3\,s (single) & 17.9 \\
				30\,s (single) & 18.9 
			\enddata
			
			\label{tab: limiting mag}
			
			
		\end{deluxetable}
		

		\begin{figure}
			
			\centering
			
			\pic[1]{lim_mag_30s}
			\caption{Same as Figure~\ref{fig: lim_mag_25Hz} only for a single, 30 second exposure. 
				The limiting magnitude from the fit is 18.9\,mag.}
			\label{fig: lim_mag_30s}
		\end{figure} 
				
		\subsection{Photometric precision}\label{sec: photometric precision}
		
		To estimate the relative error in the W-FAST data we took continuous data for 30 minutes
		at low airmass in the galactic plane at J2000 $\alpha = 19^h35^m00^s$, $\delta= 21^\circ54^{'}00^{''}$. 
		We produced lightcurves using the forced photometry pipeline described above. 
		Fluxes were background subtracted 
		and outlier measurements with flux deviating from the mean by more than $3\sigma$ were removed. 
%
		
		
		Then we correct for the zero-point (ZP), 
		or the relative transparency of the sky in each image. 
		The value of the ZP changes between images as the sky transparency changes over time, 
		and to first order we assume it affects all the stars in the image in the same way. 
		We calculate the weighted average of star fluxes, 
		where the weight for each star is the square root of the absolute value 
		of the mean flux of each star over all frames, 
		$w_i=|\sum_j f_{ij}|^{1/2}$, 
		where the index $i$ is over stars, and the index $j$ is over frames. 
		The corrected fluxes are then given by: 
		\begin{equation}
			f'_{ij} = f_{ij} / zp_j \quad,\quad zp_j = \sum_{i=1}^{N_\text{stars}} w_i f_{ij}.
		\end{equation}
		These weights were chosen after some trial and error, 
		with the final choice made to give less weight to faint stars, 
		which are more affected by certain systematics, 
		e.g., background subtraction errors, 
		and less by the sky transparency. 
		
		The zero point correction reduces much of the low-frequency variations in the lightcurves, 
		but additional improvements can be made, 
		e.g., fitting the ZP to each star's color or position in the field of view, 
		by fitting the airmass variation to each star's lightcurve, 
		or by using, algorithms designed to remove systematic effects, 
		e.g., Sys-rem \citep{sysrem_algorithm_Mazeh_2006}.
		
		After correcting for the ZP variations, 
		we calculate the relative RMS in the lightcurve of each star. 		
		The RMS can be measured for the native 25\,Hz frame rate, 
		and for a binned lightcurve, 
		where 128 measurements ($\approx 5$\,s) are averaged before the RMS is calculated.
		
\begin{figure}
			
			\centering
			
			\pic[1]{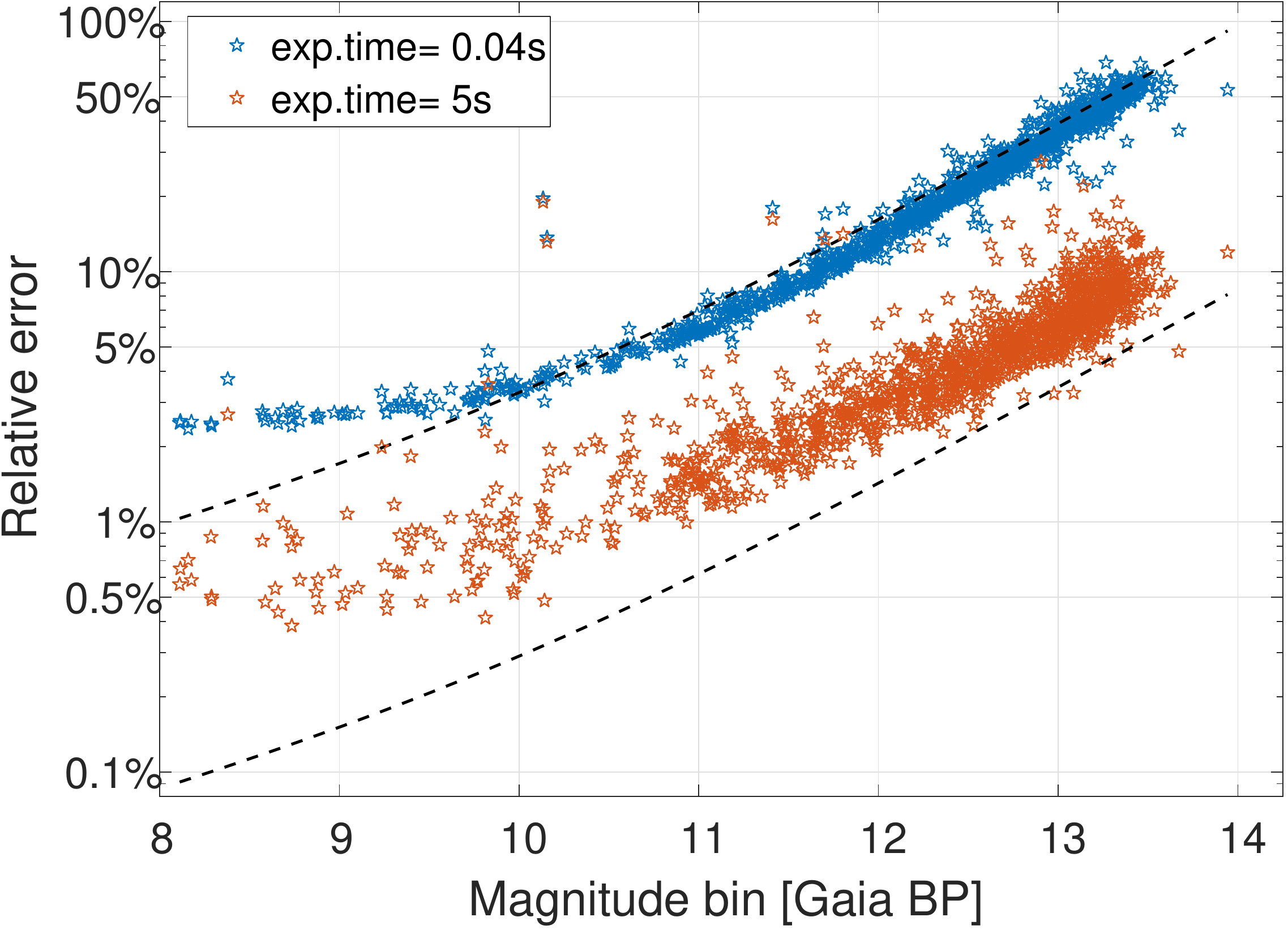}
			\caption{Relative error for the calibrated fluxes, 
				     as function of star magnitude for native-cadence (40\,ms exposures)
				     and binned lightcurves (5\,s exposures). 
					 In the binned lightcurves 
					 the faint stars gain a substantial improvement in relative RMS, 
					 since the read-noise is uncorrelated and quickly decreases when averaging.
					 The dashed lines show the theoretical noise, 
					 including a constant term (read noise and background), 
					 and a Poisson term with gain of 0.8. 
			}
			\label{fig: rms vs mag low airmass}
			
			
		\end{figure}

		In Figure~\ref{fig: rms vs mag low airmass}
		we show the RMS values for each star, in native 40\,ms and binned 5\,s cadences. 	
		The theoretical noise is shown for each time-scale. 
		The upper dashed line is calculated using a combination 
		of a Poisson term, proportional to the square root of the flux multiplied by a gain value of 0.8, 
		and constant noise term, including read noise and sky background.
		The background is estimated by the median of the background values of all stars, 
		measured using an annulus around each star. 
		
		

		The brighter stars have lower noise, as expected. 
		There are very few saturated stars in most W-FAST fields at 25\,Hz, 
		but bright stars ($\lesssim 10$ mag) no longer
		show improved relative RMS with increased flux, 
		as atmospheric scintillations and flat field errors 
		become dominant over the source noise. 
		Fainter stars have very high noise at 25\,Hz cadence, 
		but improve drastically as the binning reduces the uncorrelated read-noise. 
		At 5s intervals, the error ranges between 0.5 and 5\%. 
		
%
%
%
%
%
%
%
%
%
%

		
	\section{Main science goals}\label{sec: science}
		
		W-FAST is built to explore the phase space of 
		wide field-of-view and high cadence. 
		This includes periodic and variable sources, 
		as well as short-duration transients. 
		We review some of the science goals in the following subsections. 
		
		\subsection{Solar System occultations}
		
		One of the science goals of W-FAST is to use occultations of background stars
		to detect Solar System objects that are too faint to detect in reflected light \citep{occultation_idea_Bailey_1976}.
		The first objects we expect to detect using this method are Kuiper Belt Objects, 
		which are estimated to be detectable at a rate of $\approx 1$ per month (see below).  
		Another class of objects are Oort cloud objects for which the detection rates are uncertain. 
		
		The Kuiper belt is interesting as it presumably represents a pristine population
		of remnant comets left over from the early stages of Solar System formation. 
		Particularly the smaller, km-scale KBOs can be used to better understand the processes of planetary formation
		\citep{KBO_sculpting_Morbidelli_Brown_Levison_2003,KBO_initial_sizes_Schlichting_2013}. 
		Measuring the size distribution of small KBOs provides information about the collisional history of the Kuiper belt, 
		and can constrain the composition and material strength of these objects \citep{KBOs_power_law_Pan_Sari_2004}, 
		while the inclination distribution of small KBOs may provide clues to their origins and interaction history. 
		Furthermore, understanding the rate of collisions (through measuring the size and velocity distributions), 
		will provide an important measurement of the dust production rate in the outer Solar System. 
		Comparing this rate to the rate of dust production in debris disks around other stars may provide a way to 
		estimate the size and extent of the parent populations (exo-Kuiper belts). 
		
		
		Even though small KBOs, of order 1\,km radius, are too faint to detect directly, 
		\cite{KBO_single_object_Schlichting_Ofek_2009,abundance_kuiper_objects_Schlichting_Ofek_2012}
		demonstrated that they can be detected by their occultation of background stars.
		Recently, a km-scale KBO has also been detected by a ground based observatory \citep{KBO_detection_from_ground_Arimatsu_2019}.
		
		The statistics given by previous detections of small KBOs \citep{KBO_single_object_Schlichting_Ofek_2009,abundance_kuiper_objects_Schlichting_Ofek_2012}
		imply that an occultation of a single star by a 0.5\,km radius KBO occurs, on average, every $\sim 10^4$ hours. 
		For this reason, W-FAST is designed to photometrically follow thousands of stars in the field at a cadence of 10--25\,Hz. 
		
		An even more ambitious goal of W-FAST is detection of the Oort cloud \citep{Oort_cloud_hypothesis_Oort_1950}.
		To date there has been no in-situ detection of any member of the Oort cloud,
		so the actual number, size and distance distributions 
		of the Oort cloud objects are still poorly constrained. 
		Models of the size and distance distribution of the Oort cloud,
		based on dynamical simulations, 
		predict an order of $10^{12}$ comets larger than 1\,km in the inner regions of the Oort cloud 
		\citep{oort_cloud_formation_Duncan_1987,oort_cloud_origin_size_Morbidelli_2005,oort_cloud_flux_comets_Emelyanenko_2007}, 
		but these models have large uncertainties regarding the number and distance distributions. 
		An occultation survey is the most promising way we know of to try to detect the Oort cloud, 
		even though it is not known \emph{a priory} how much observation time 
		will be required for a single detection. 
		As with the Kuiper belt, measuring the properties of the Oort cloud 
		can help constrain models of Solar System formation and may help 
		us to narrow the search for equivalent clouds around other stars (e.g., \citealt{debris_disks_Herschel_Dodson_Robinson_2016}).
		It may also be useful for constraining the expected flux of comets expelled from other stars, e.g., \cite{interstellar_comet_2I_Borisov_Jewit_Luu_2019}. 
		Discovering even a single object, 
		or placing strong constraints on the population of the inner Oort cloud, 
		would greatly advance our understanding of comet production and ejection in the early Solar System. 
		

		\subsection{Accreting black holes}
		
		Accreting stellar-mass black holes 
		can be used to study accretion processes on short time-scales 
		that can illuminate the physics of larger systems like Active Galactic Nuclei (AGN).
		Understanding the accretion of matter in extreme gravity is relevant 
		to study of scaled-up versions of these systems, 
		e.g., superluminous X-ray sources \citep{ultraluminous_x_ray_sources_compared_to_stellar_black_holes_Miller_2004,ultraluminous_x_ray_source_optical_counterpart_Kuntz_2005}.
		Stellar mass black holes are presumably remnants of core-collapse supernovae 
		so that measuring their mass distribution, binary orbital parameters and space velocities,
		can help constrain the possible types of explosions we see in massive stars \citep{x_ray_binaries_supernovae_connection_Casares_2017}. 
		They are detected almost exclusively as the heavier object in binary systems, 
		often very close binaries that can only come to exist through some form of binary evolution, 
		thus their orbital properties provide clues about the interaction history of the system. 
		Finally, these binaries eventually continue to evolve to form binaries of two compact objects (neutron stars or black holes),
		of which some are the progenitors of gravitational waves that are detected by LIGO/Virgo (e.g., \citealt{gravitational_waves_first_detection_Abbott_2016}). 
		
		As gas falls towards the black hole it needs to shed angular momentum and 
		presumably forms an accretion disc. 
		The accretion disc itself contributes to the thermal component of the emission, 
		with the luminosity and peak energy depending on the accretion rate and disc inner radius \citep{x_ray_binaries_black_hole_review_Remillard_McClintock_2006}. 
		Along with the disc, a corona (optically thin material around the compact object or above the disc)
		and possibly a jet can contribute to the hard component of the spectrum \citep{black_holes_compact_review_Bambi_2018}.
		The relative luminosity of these components can change with time, 
		possibly due to the disc transitioning between different states, 
		so the objects occasionally move from one spectral state to another. 
		Studying the changes between spectral states, 
		as well as the differences between particular objects that
		have different masses, accretion rates and inclinations, 
		sheds light on the physics and geometry of the different components. 
		
		There are only $\approx 20$ confirmed accreting black hole binaries
		in our galaxy and in neighboring galaxies, 
		and $\approx 50$ more that are black hole candidates for which dynamical mass 
		could not be inferred \citep{accreting_black_holes_BlackCAT_Corral-Santana_2016}. 
		Some of these systems are persistent X-ray sources, usually the High Mass X-ray Binaries (HMXB), 
		while many more are transient: almost all the Low Mass X-ray Binaries (LMXB). 
		One or two transient systems are detected each year in outburst, 
		when they are brightest in the X-ray bands. 
		It is estimated that there are 
		of order $10^3-10^4$ black hole LMXBs in the galaxy 
		\citep{black_hole_x_ray_binary_population_Tanaka_1996,black_hole_x_ray_binary_population_Yungelson_2006,accreting_black_holes_BlackCAT_Corral-Santana_2016}, 
		most of which are transients that have yet to go into outburst 
		since X-ray satellites were first launched. 
		The low duty cycle of transient LMXBs, combined with 
		the lack of any detected eclipsing black hole binaries ~\citep{black_hole_x_ray_binary_inclinations_Narayan_McClintock_2005}
		suggests that many more such systems can be found if the selection 
		of objects is not made only in X-rays, 
		e.g., recently discovered MAXI \mbox{J1847-004}, \citep{x_ray_source_MAXI_new_transient__ATEL_Negoro_2019}, 
		was undetected until the recent outburst. 
		
		While most binary black holes are detected by their X-ray emission, 
		it may also be possible to detect them by other means. 
		As an example, VLA J213002.08+120904 \citep{black_hole_x_ray_binary_detected_quiescence_Tetarenko_2016}
		is the first field black hole LMXB detected in quiescence, 
		by observing its radio emission. 
		Another black hole candidate system, AT2019wey \citep{x_ray_binary_optical_detection_Yao_2020,x_ray_binary_resolved_source_Yadlapalli_2020}
		was detected as an optical transient before its X-ray properties were discovered. 
		It may also be possible to detect such systems based on their 
		visible light, either by their variability or periodicity. 
		Optical oscillations in the quiescent optical emission may be caused by 
		elliptical modulation of the donor star \citep{low_mass_x_ray_binaries_multiwavelength_review_Hynes_2010}, 
		by warping of the accretion disc (i.e., ``superhumps'', \citealt{black_hole_x_ray_binary_superhumps_ODonoghue_1996,black_hole_x_ray_binary_superhump_telegram_Uemura_2000,black_hole_x_ray_binary_quiescent_superhumps_Zurita_2002,black_hole_x_ray_binary_superhumps_Neil_2007}) 
		or by other mechanisms causing Quasi Periodic Oscillations (QPOs, e.g., \citealt{black_hole_x_ray_binary_quasi_periodic_oscillations_Motta_2016}).
		An all-sky survey in the optical, with good photometric stability, may be sensitive to these oscillations  
		(e.g., as suggested by \citealt{black_hole_search_LAMOST_Yi_2019,black_hole_search_LAMOST_ASASSN_Zheng_2019}).
		A few systems may be close enough to be detected in outburst, 
		but are seen edge-on so their X-ray emission is hidden. 
		Such systems also demonstrate periodicity or variability
		e.g., by irradiance of the donor star, 
		without being detected by X-ray surveys. 	
		
		Detecting black hole binaries in the optical is challenging. 
		However, even a handful of new systems 
		can significantly increase the census of galactic black holes, 
		especially if some of them are edge on systems, 
		that are not easily detected by X-rays. 
		
		
		Furthermore, the optical outbursts often precede the X-ray outburst, 
		as the increased accretion rate moves from the outer disc inwards.
		So an optical monitoring program that detects 
		the optical outburst as it begins may leave enough time 
		to point X-ray telescope at a source \emph{before} the X-ray outburst begins,
		as suggested by \cite{black_hole_x_ray_binary_optical_precursors_Russell_2019}. 
		Finally, other types of transients that have not yet been thoroughly studied 
		because their outburst is too soft for detection by X-ray surveys 
		may be detectable in optical surveys that look for very fast variability, 
		e.g, the mysterious optical/X-ray counterpart to GRB070610, 
		which is likely a galactic source \citep{x_ray_binary_GRB_transient_Kasliwal_2008}.

		W-FAST will perform a galactic plane survey, 
		searching for variability on different time scales. 
		Black hole binary candidates can be detected by 
		their specific periodic signal, 
		for which W-FAST's continuous coverage is particularly useful 
		as it helps avoid aliasing with the base period. 
		W-FAST may also be sensitive to short time-scale flickering,
		caused by non-uniformities in the accretion flow. 
		
		There are eight low mass black hole binaries (confirmed and candidates) within a distance of $\lesssim 3$\,kpc. 
		Assuming the duty cycle of these systems is $\lesssim 10$\%, 
		we would expect to have an order of 100 systems that have not yet been detected in outburst, 
		and assuming their spatial distribution is along the galactic plane, 
		we should be able to see about ten new systems within 1\,kpc, 
		where a low mass companion can be detected and its variability measured.
		For W-FAST at high-cadence mode using stacks of 100 images, 
		we would be able to detect K0 stars or brighter. 
		Using native 3\,s exposures, 
		we could detect M0 stars, with up to 1 magnitude of extinction, 
		up to such a distance. 
				
				
		\subsection{Spinning magnetic white dwarfs}
		
		A more abundant class of objects that also allow us to investigate 
		accretion physics and binary evolution are the Cataclysmic Variables (CVs), 
		where the accretion is onto a white dwarf \citep{cataclysmic_variables_review_Giovannelli_2008}. 
		The flux is typically dominated by the accretion disc, 
		and there are no additional components such as a jet or a corona, 
		and very little contribution of irradiation of the disc, 
		so these systems are also cleaner laboratories for study of the accretion disc physics. 
		White dwarf binaries are bright in the optical, 
		abundant and nearby, making them easier to study than X-ray binaries. 
		Binaries including white dwarfs are important in understanding binary stellar evolution 
		and are possibly the progenitors of type Ia supernovae (either as single- or double-degenerate systems).

		A subset of CVs include white dwarfs with strong magnetic fields \citep{cataclysmic_varaibles_magnetic_review_Wickramasinghe_Ferrario_2000}. 
		In the case of AM-Her stars (polar CVs) the accretion is completely dominated 
		by the magnetic field of the white dwarf, 
		and the material never accumulates into a disc. 
		In these systems the white dwarf spin is synchronized to the binary period. 
		In DQ-Her stars (intermediate polars, IPs), 
		the magnetic field only partially disrupts the accretion disc, 
		so its inner portion is replaced by accretion along the magnetic field lines, 
		directly onto the surface \citep{intermediate_polars_review_Patterson_1994}. 
		In these systems the white dwarf rotates faster than the binary period, 
		and in some cases the period can be very short 
		(e.g., in DQ-Her the white dwarf has a spin period of 71 seconds; see Section~\ref{sec: dq her}). 		
		As the white dwarf rotates inside the accretion disc, 
		its magnetic poles can illuminate the disc and cause 
		short period modulations. 
		While IPs are usually found in outburst, 
		they can also be detected in quiescence by looking for these fast oscillations. 
		
		It is useful to be able to find non-erupting magnetic CVs. 
		One hypothesis that has not yet been confirmed, 
		is that IPs evolve into polar CVs 
		(e.g., \citealt{intermediate_polar_EI_UMa_Reimer_2008}). 
		This is consistent with current observations, 
		however, a better census of the number of systems 
		of each type can be used to verify or refute this idea. 
		Furthermore, magnetic CVs are thought to dominate the diffuse hard X-ray flux 
		in the bulge of the galaxy \citep{cataclysmic_variable_space_density_Pretorius_2014, cataclysmic_variables_intermediate_polars_space_density_Pretorius_Mukai_2014}, 
		so improving the census of the population of the faintest IPs, 
		that are likely to be more numerous than their bright counterparts, 
		is an important way to account for the diffuse flux. 
		Lastly, magnetic white dwarf binaries are natural laboratories for 
		studying physics of plasma in strong magnetic fields. 
		There are a few tens of confirmed 
		IPs\footnote{https://asd.gsfc.nasa.gov/Koji.Mukai/iphome/catalog/alpha.html}, 
		so even a few more tens of non-erupting such systems could contribute 
		to our understanding of the population statistics, 
		e.g., low accretion systems near the 80 minute period limit. 
		
		W-FAST will have the advantage of high-cadence, continuous coverage
		when searching for short-time periodic signals. 
		Even low amplitude periodic variations could be detected
		when added coherently on a long enough time-span. 

	\section{First results}\label{sec: first results}
		
		\subsection{Targeted KBO occultation}
		
		On the first night of commissioning we observed 
		an occultation by a known KBO, 
		38628 Huya, which is a $\approx 400$\,km object in the Kuiper belt. 
		The occultation was known in advance, and we could target the field 
		ahead of time, unlike in our normal un-targeted, small KBO survey. 
		
		During the first nights of commissioning,
		on 2019 March 18, we took images at one second cadence of a field 
		at $\alpha=16^{h}41^{m}06.4^{s}$, $\delta=-06^\circ43^{'}34.6^{''}$ (J2000), 
		using a prototype camera (The Andor Zyla 5.5) that has smaller pixels and a smaller field of view.
		We generated lightcurves for 500 stars using a 5 pixel ($6.25"$) aperture, 
		and an 8-11 pixel ($9-13.75''$) annulus for background subtraction. 
		Both aperture and annulus were shifted to the position 
		of the center of light for each star in each image.
		%
		%
		To correct for varying sky conditions 
		we fit a second order polynomial to each lightcurve, 
		iteratively removing $3\sigma$ outliers. 
		
		%
		%
		%
		%
		%
		%
		%
		
\begin{figure}
			
			\centering
			\pic[\picsize]{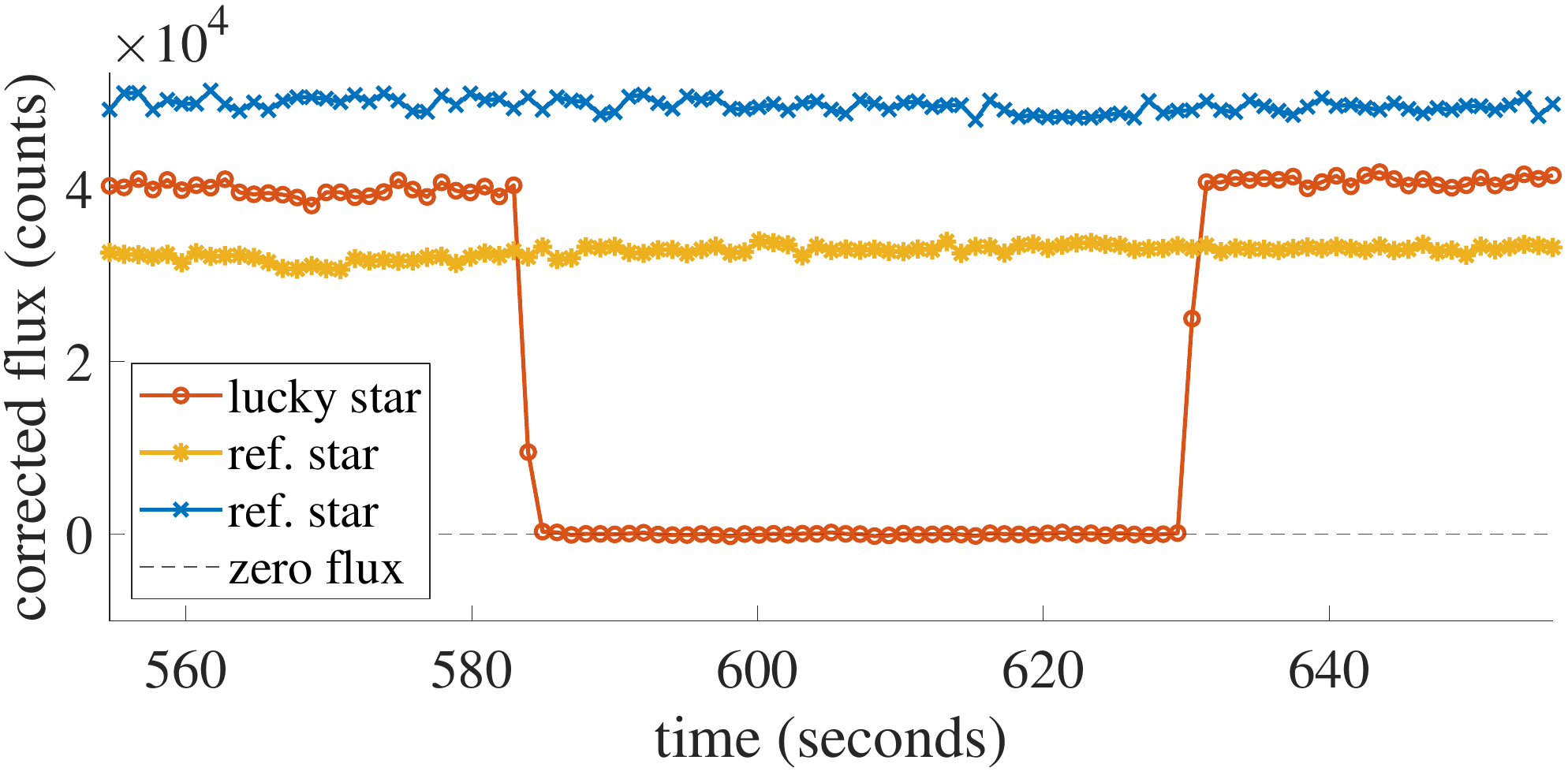}
			\caption{Lightcurves of 18th, 21st and 28th brightest stars in the field, 
				showing the occultation by Huya. 
				The reference stars plotted here show the instrument did not suffer 
				any major loss of light for other stars in the field. 
				The timing and length of the occultation is consistent with other observatories 
				viewing the same event \citep{transneptunian_object_Huya_occultation_Santos_Sanz_2019}. 
				The light of the occulted star is completely dimmed, 
				and the data points at the edge of the occultation show the timescale for entering 
				and exiting the shadow is about one second. 
				The entire occultation lasted 44.4s.}
			
			\label{fig: huya closeup}
			
		\end{figure}

		The star that was occulted by Huya has a magnitude of $m_V=10.4$.
		In Figure~\ref{fig: huya closeup} we show the occulted stars
		with two other stars of similar brightness.
		The photometry for other stars seems to be stable at this time.
		The occultation occurred between 00:25:34 -- 00:26:18 UTC, lasting 44.4s, 
		with one data point during the entrance and one during the exit from the shadow. 
		The object's sky velocity\footnote{https://lesia.obspm.fr/lucky-star/occ.php?p=15369} was 8.2\,km\,s$^{-1}$, 
		indicating the size of the KBO is at least 364\,km. 
		Since the observatory did not necessarily cross the center of the shadow, 
		this is only a lower bound on the diameter of Huya,
		which is estimated to be above 400\,km. 
		
		
		\subsection{Cataclysmic variables: DQ Her}\label{sec: dq her}
		
		DQ Her is an intermediate polar, 
		with a white dwarf rotating on a shorter period 
		than the orbital period.
		The orbital period is 4.65 hr 
		while the rotation period of the white dwarf is 71s,
		and the $B$-band magnitude of the system is around 14.5
		\citep{cataclysmic_variables_rotation_DQ_Her_Patterson_1978}.

\begin{figure}
			
			\centering
			
			\pic[1]{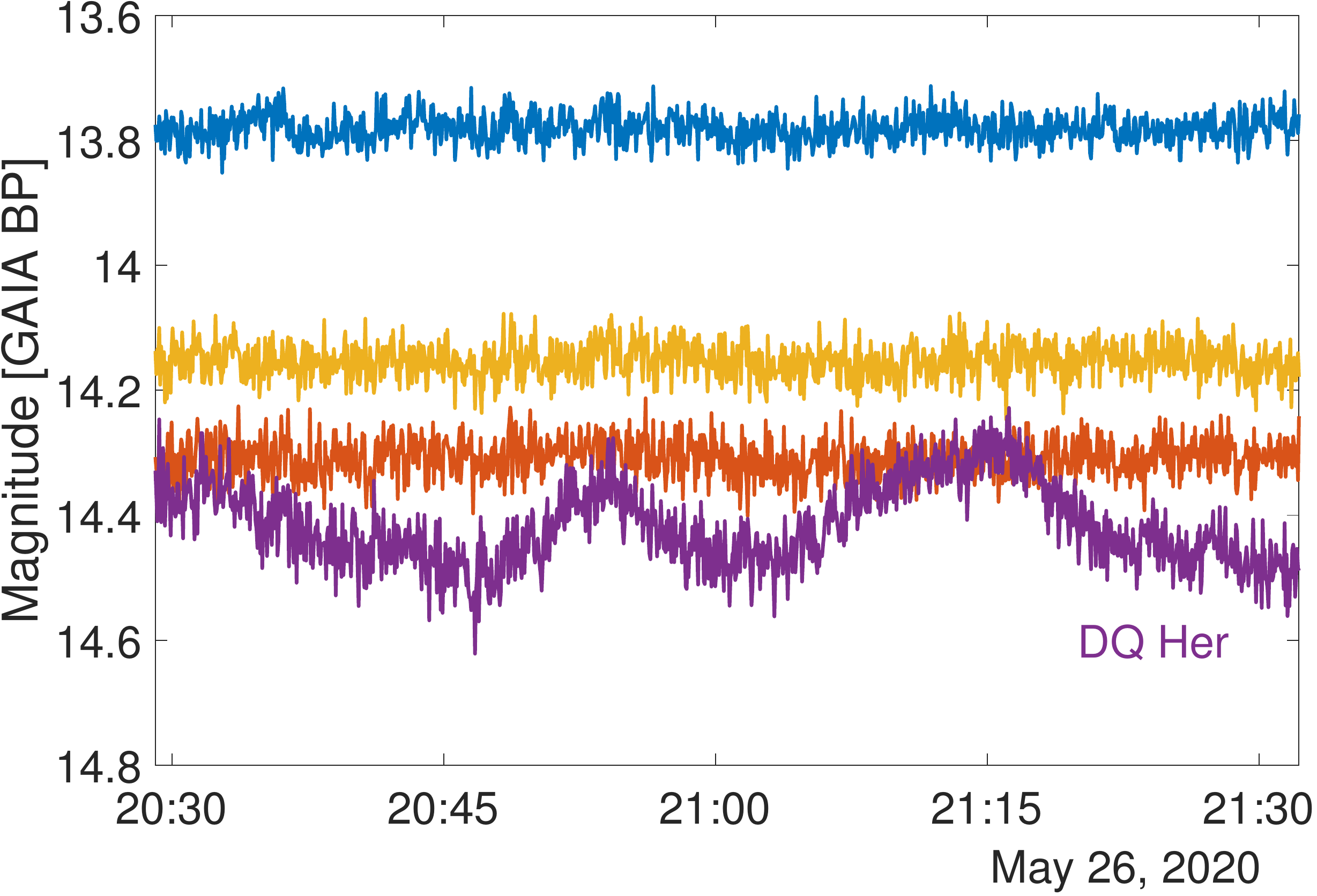}
			
			\caption{Lightcurves of DQ Herculis and three other stars in the field, 
				     with similar magnitudes, for comparison.
				     Data was taken at cadence of about 3s. 
			         The lower, purple plot is for DQ Her. 
			         The relative RMS of the other stars is around 2-3\%, 
			         while the DQ Her lightcurve shows variability of 6\%. 
			         The apparent variations are not periodic, 
			         but a 71s period is buried under the noise. 
		          }
			\label{fig: dq her lightcurves}
			
			
		\end{figure}

		We obtained an hour of observations at 2020 May 26, 
		of a field centered on DQ Herculis. 
		We used slow-mode (native 3s exposure times) 
		to observe this target.
		We show the resulting lightcurves for DQ Her 
		and a few other stars of similar magnitudes in Figure~\ref{fig: dq her lightcurves}. 
		The lightcurves of the other stars, 
		shown here for comparison, 
		have a relative RMS of about 2-3\%. 
		The lower plot, for DQ Her, 
		shows variability of 6\%, 
		with obvious structure, 
		but no apparent periodic signal. 
		When plotting the power spectrum 
		we see a clear peak at a period of 71.4s. 
		This is consistent with the white dwarf rotation period. 
		The power at lower frequencies is due to 
		systematic red-noise and shot-noise from the object itself. 
		
		\begin{figure}
			
			\centering
			
			\pic[1]{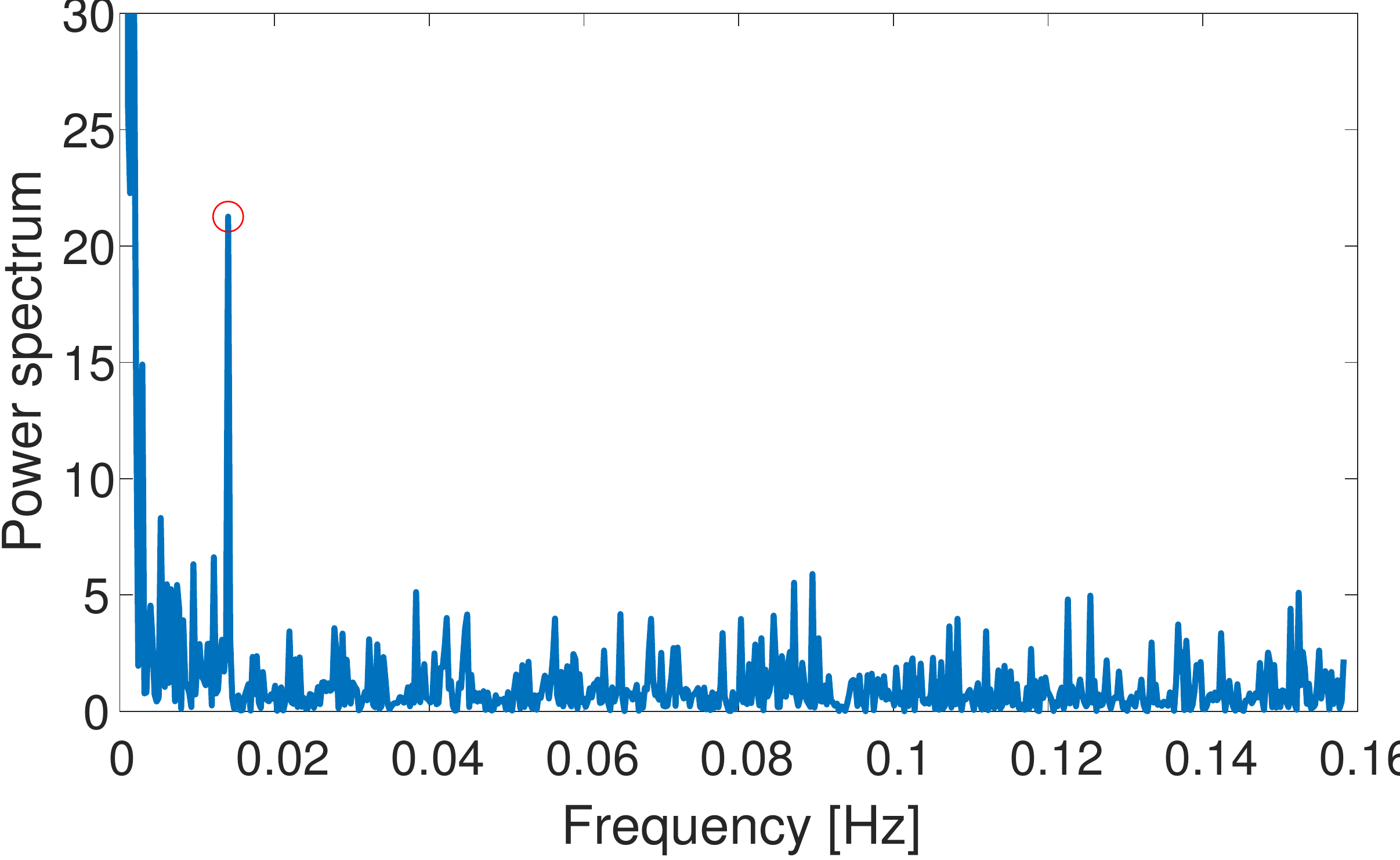}
			
			\caption{Power spectrum of DQ Her, taken at 3s cadence. 
			         The power in the lower frequencies represents red noise from various sources. 
			         The peak marked with a red circle is for the period of 71.4s, 
			         consistent with the white dwarf rotation period of 71s. 
			         The plot is normalized by the standard deviation of 
			         the power spectrum above 0.02\,Hz. 
		         }
			\label{fig: dq her ps}
			
			
		\end{figure}

	\subsection{Search for fast transients}
	
	During 2020 July and August we conducted a blind search for 
	sub-second transients using a custom pipeline running
	that search for transients in the 25\,Hz, full-frame imaging data. 
	We have detected a high rate (30--40 events per day per deg$^2$) 
	of short-duration glints (on the order of 0.2\,s), 
	with magnitude in the 9--11 range, 
	coming from geosynchronous satellites
	\citep{satellite_glints_WFAST_Nir_2020,satellite_glints_GN_z11_flash_Nir_2021}.

	These glints would be an important foreground
	to any searches for astrophysical transients. 
	We plan to conduct a survey targeting the Earth's shadow, 
	where satellites cannot glint, 
	to better constrain the rate
	of short duration astrophysical events. 
		
	\section{Summary}\label{sec: summary}
	
	The W-FAST system is a 55\,cm, f/1.9 optical observatory
	that can observe a field of view of 7\,deg$^2$ at a rate of up to 25\,Hz. 
	This system is used to explore 
	a new parameter space of fast cadence and large field of view. 
	It can be used for detecting occultations by KBOs 
	and possibly other, further Solar System objects. 
	

	The short cadence will make it possible to 
	uniquely identify short period objects without aliasing, 
	as well as allow detection of even faster phenomena
	on second or sub-second time-scales. 
	
	We show the system can reach a photometric precision
	of better than 1\% for bright stars ($B_p<10$) 
	and 2\% for faint stars ($B_p<13$) 
	when summing data from multiple exposures at $\sim$minute time-scales. 
	With a large sample of stars in each field, 
	self-calibration could help reduce some of the systematics
	involved with co-adding images over longer time-scales, 
	reaching down to few milli-mag precision. 	
	Moreover, combining multiple short exposures into longer, deeper images 
	yields some advantages over taking natively long exposures, 
	when dealing with tracking errors, cosmic rays, and systematic red noise. 
	
	In the near future we intend to conduct a galactic plane survey, 
	to characterize stellar variability on second time-scales
	and to detect short period binaries. 
	We are also actively searching for astrophysical sub-second transients 
	inside the Earth's shadow, where flashes from high-orbit satellites
	should not be visible \citep{satellite_glints_WFAST_Nir_2020}.
	
	\section*{Acknowledgments}
	
	We are thankful to all the staff and students that helped
	build the observatory: 
	I.~Bar,
	S.~Ben-Gigi,
	R.~Bruch,
	B.~Callendret,
	T.~Engel,
	A.~Ghosh,
	S.~Goldwasser,
	I.~Irani,
	S.~Kaspi,
	D.~Khazov,
	N.~Knezevic,
	S.~Knezevic,
	A.~Lahmi, 
	J.~Mushkin,
	B.~Pasmantirer,
	M.~Rappaport,
	B.~Reyna,
	A.~Rubin,
	H.~Sade,
	I.~Sagiv, 
	Y.~Shachar,
	A.~Sharon,
	M.~Soumagnac,
	N.~Strotjohann.
	E.O.O.~is grateful for the support by grants from the 
	Israel Science Foundation, Minerva, Israeli Ministry of Science, Weizmann-UK, 
	the US-Israel Binational Science Foundation,
	and the I-CORE Program of the Planning
	and Budgeting Committee and The Israel Science Foundation.
	AGY’s research is supported by the EU via ERC grant No. 725161, the ISF GW excellence center, an IMOS space infrastructure grant and BSF/Transformative and GIF grants, as well as The Benoziyo Endowment Fund for the Advancement of Science, the Deloro Institute for Advanced Research in Space and Optics, The Veronika A. Rabl Physics Discretionary Fund, Minerva, Yeda-Sela and the Schwartz/Reisman Collaborative Science Program;  AGY is the recipient of the Helen and Martin Kimmel Award for Innovative Investigation.
	B.Z.~is supported by a research grant from the Ruth and Herman Albert Scholarship Program for New Scientists.
	We thank the Wise Observatory staff for their continuous support.

	
	\bibliographystyle{aasjournal}
	
	\bibliography{refs}
	
\end{document}